\documentclass[fleqn,usenatbib]{mnras}

\usepackage[T1]{fontenc}
\usepackage{ae,aecompl}

\usepackage{graphicx}	
\usepackage{amsmath}	
\usepackage{amssymb}	

\usepackage{newtxtext,newtxmath}

\usepackage{etoolbox}

\AtBeginShipout{%
  \ifnum\value{page}>2 %
    \typeout{* Additional boxing of page `\thepage'}%
    \setbox\AtBeginShipoutBox=\hbox{\copy\AtBeginShipoutBox}%
  \fi
}

\usepackage[table, dvipsnames]{xcolor}

\title[Effect of {\it{{\it{Gaia}}}}-based data on grid search models]
{Uniform characterisation of an ensemble of main-sequence benchmark stars: effect of {\it{{\it{Gaia}}}}-based data on grid search models}
\author[Nsamba et al.]{
Benard Nsamba, $^{1,2,3}$\thanks{E-mail: nsamba@mpa-garching.mpg.de}
Achim Weiss, $^{1}$ and Juma Kamulali, $^{2}$
\\
$^{1}$ Max-Planck-Institut f\"{u}r Astrophysik, Karl-Schwarzschild-Str. 1, D-85748 Garching, Germany\\
$^{2}$Department of Physics, Faculty of Science, Kyambogo University, P.O. Box 1, Kyambogo, Kampala, Uganda\\
$^{3}$Instituto de Astrof\'{\i}sica e Ci\^{e}ncias do Espa\c{c}o, Universidade do Porto,  Rua das Estrelas, PT4150-762 Porto, Portugal\\
}

\date{Accepted 2024 December 05. Received 2024 November 23; in original form 2023 December 19}

\pubyear{2024}

\begin{document}
\label{firstpage}
\pagerange{\pageref{firstpage}--\pageref{lastpage}}
\maketitle
\begin{abstract}
The inference of stellar parameters (such as radius and mass) through asteroseismic forward modelling depends on the number, accuracy, and precision of seismic and atmospheric constraints. 
%
ESA's {\bf{{\it{{\it{Gaia}}}}}} space mission is providing precise parallaxes which yield an additional constraint to be included in the model grid search. Using a handful of main-sequence benchmark stars, we perform a uniform characterisation of these stars. We assess the accuracy and precision of stellar parameters inferred from grid-based searches when a {\bf{{\it{{\it{Gaia}}}}}}-based luminosity is combined with different stellar constraints.
We also examine the  precision needed for an interferometric radius (model-independent radius) to have a significant contribution towards the determination of stellar mass in the optimisation process.
Our findings show that more precise stellar masses are inferred for some stars
when seismic and spectroscopic constraints are complemented with a Gaia-based luminosity, with a scatter varying from 1.9 per cent to 0.8 per cent. However, the inferred stellar radii are underestimated when compared to the interferometric radii and yield a scatter of $\sim$1.9 per cent. 
In addition, we demonstrate that a precisely measured interferometric radius ($\lesssim$ 1 per cent) when applied in the optimisation process yields a mass with a precision $\lesssim$ 1.5 per cent. 
Finally, we find that when only $l=0$ mode oscillation frequencies are available, robust masses and radii are still  attainable. However, this requires  precise and numerous $l=0$ mode oscillations frequencies ($>$ 8) to be coupled with atmospheric constraints.

\end{abstract}

\section{Introduction}
Asteroseismology is a powerful tool used to characterise stellar interior structures, test the understanding and description of stellar physics, and infer fundamental stellar properties such as the mean density, radius, mass, and age, e.g. \citet{2005Miglio,2010Metcalfe,2011Aguirre,2012Creevey,2013Do,Metcalfe_2014,2014Valle,2015Valle,2015Aguirr,Davies,2018MNRANsamba,2018MNsamba,2018ADeal,10Nsamba,2019Valle,2020Moedas,2020Deal}. The knowledge of the fundamental stellar properties has been employed in making essential contributions towards the understanding of other astrophysics research fields such as Galactic archaeology (e.g. \citealt{2017Miglio,2023ner}) and Exoplanet studies (e.g. \citealt{2009Huber,Benomar9,2014Lebreton,Metcalfe_2014,2016Lundkvist,Campante_2016,2019Campante,2019Huber,2020Jiang,2020Nielsen}, just to mention a few). 

The continuous advancement of asteroseismiology research is greatly attributed to a handful of space missions such as the French led CoRoT (Convection, Rotation and Planetary Transits; \citealt{2009Auvergne})  mission, NASA's {\it Kepler} space mission \citep{Borucki977}, NASA's TESS (Transiting Exoplanet Survey Satellite; \citealt{George}) mission, and with the future ESA PLATO (PLAnetary Transits and Oscillations of stars; \citealt{Rauer2014}) mission. This is partly because space missions provide uninterrupted photometric data with significantly minimised noise levels, yielding a frequency spectrum with a range of excited modes which can be easily identified (see \citealt{2004Kjeldsen,2005Kjeldsen}). 
The exceptional photometric observations  have given rise to continuous development of asteroseismic tools which have been employed successfully to infer precise  star parameters like radius, mass, and age (e.g. \citealt{2017Bellinger,2021Salmon,2023Guo}). The systematic uncertainties (scatter) on stellar parameters derived using different seismic optimisation tools have also been explored (see \citealt{2009Monteiro}, and \citealt{2015Aguirr,2017Aguirre}). Furthermore, extensive exploration of the impact of model physics such as atomic diffusion including gravitational settling \citep{2014Valle,2015Valle,2018MNRANsamba}, radiative acceleration \citep{2018ADeal,2020Moedas,2020Deal}, solar metallicity mixtures \citep{2018MNsamba,10Nsamba,2019Valle}, 
opacities \citep{Huebner2014}, equation of states \citep{1998Rogers}, rotational mixing, semi-convection, core and envelope convective overshooting \citep{2005Miglio,2011Aguirre,2012Creevey,2015Valle,2022Ahlborn}, 
on the derived stellar parameters have also been a center of attention for about a decade or so. All these efforts have generated a leap forward in our understanding of additional uncertainties to be considered on stellar parameters derived using asteroseismology.

Fundamental stellar parameters inferred using asteroseismology are commonly based on ``Forward modelling techniques''. This involves building dedicated internal structure models which represent the targeted/observed star. This is also known as ``a 'la carte modelling'' \citep{2014Lebreton}. Based on a set of available observational stellar data, an optimisation process is carried out to determine the stellar model which best matches the available data. There are two sets of observational data usually considered in the optimisation process, i.e.\ seismic data which includes global oscillation parameters (i.e.\ frequency of maximum power, $\nu_{\rm{max}}$, and large frequency separation, $\Delta \nu$), individual oscillation frequencies, $\nu_{\rm i}$, or their combinations \citep{2003Roxburgh}, and classical constraints composed of effective temperature, $T_{\rm{eff}}$, metallicity, [Fe/H] or individual abundances, luminosity, $L$, and when available a model-independent radius, $R$.
Although this approach is known to yield precise stellar parameters, one has to keep in mind that the inferred stellar parameters are model-dependent and therefore sensitive to the input physics used in the models (e.g. \citealt{2015Aguirr, 2018MNRANsamba, 2022Moedas}).

With the tremendous success of asteroseismic techniques towards the determination of stellar parameters, it is relevant to establish the accuracy and precision of asteroseismic inferences. \citet{2017Aguirre} explored the robustness of asteroseismically determined stellar parameters for a sample of 66 {\it{Kepler}} Legacy stars whose  individual oscillation frequencies are available. 
They demonstrated that the uncertainties on derived mass, radius, and age are reduced when individual oscillation frequencies are taken into account compared to when only seismic global parameters (i.e.\ frequency of maximum power, $\nu_{\rm{max}}$ and large frequency separation, $\Delta\nu$) are considered \citep{2010Gilliland}, i.e.\ from 2.2 to 2.0 per cent in radius, 5.4 to 4.0 per cent in mass,  and 25 to 10 per cent in age, respectively.
However, this improvement in the precision of asteroseismically determined stellar parameters also necessitates corresponding tests on accuracy of these parameters. These tests can only be achieved when compared to corresponding model-independent parameters with uncertainties (both statistical and systematic) smaller than the asteroseismically determined stellar parameters \citep{2021Cunha}.

Long-baseline interferometry is reported to be successful in estimating angular diameters of bright stars, from which their corresponding radii are derived (e.g. \citealt{2003Kervella,2009Mazumdar,2012MHuber,2012Creevey,2015Creevey,2016Pourbaix,2016Lagrange,2017Kervella,2020Karovicova}). This has been used to test the precision and accuracy of stellar radii inferred using asteroseismology (e.g. \citealt{2012MHuber,2012Hub,2013White,2015EPJWC}). For instance, through the combination of {\it{Hipparcos}} parallaxes with angular diameters of five main-sequence stars, \citet{2012MHuber} calculated model-independent linear radii of these stars and compared them with asteroseismic radii (i.e.\ based on asteroseismic scaling relations for the frequency of maximum power, $\nu _{\rm{max}}$, and the large frequency separation, $\Delta\nu$). They report the asteroseismic radii to be accurate to better than 4 per cent. In addition, they anticipate improvement in the accuracy of asteroseismic radii when determined through detailed modelling involving individual oscillation frequencies.

To quantify the accuracy of seismic masses, model-independent masses are needed. Stars in binary systems offer a great opportunity of availing model-independent masses if their orbital parameters are known. The limitation to performing this exercise is that only a handful of main-sequence benchmark stars are available with both seismic and model-independent parameters, like radius and mass. To this end, previous efforts concentrated on studies of double-lined spectroscopic binaries and eclipsing binaries (e.g. \citealt{2010Torres,2020Halbwachs,2021Serenelli,2023Beck}). 
In the context of exploring the robustness of stellar parameters expected from the future ESA’s PLATO 
mission, \citet{2021Cunha} performed a ``hare-and-hounds exercise'' which involved inferring stellar parameters of six simulated (artificial) main-sequence stars. 
They reported a difference in the accuracy between simulated and ``true values'' in radius, mass, and age to be 1.33, 4.32, and 11.25 per cent, respectively. 
We highlight that the simulated models may already have suffered from incorrect physics that would reduce the accuracy of models of real stars even further.
In addition, \citet{2021Cunha} also examined how the precision and accuracy of inferred parameters would vary when different surface corrections routines and classical combinations are adopted. 
\citet{2007Creevey} used simulated data and explored the relevance of a model-independent radius and its combination with other stellar constraints (seismic and classical) towards the determination of a precise and accurate stellar mass. Their findings illustrate that a model-independent radius when included in the optimisation process plays a significant role towards the determination of a precise and accurate stellar mass.

ESA's {\it{Gaia}} provides unprecedented quantity and quality of a uniform, homogeneous, and precise data set for more than a billion stars, yielding their accurate astrometric data \citep{2018Collaboration,2023Colla}. 
\citet{2017Hub} highlights an insightful glimpse of the powerful synergy between {\it{{\it{Gaia}}}} and asteroseismology. Using 2200 {\it{Kepler}} stars at different evolution phases (i.e. from  the main sequence to red-giant branch), they presented a comparison  of radii based on {\it{{\it{Gaia}}}} DR1 parallaxes and asteroseismic scaling relations, --- thus demonstrating that asteroseismic radii determined using scaling relations are accurate to $\sim$5 per cent for stars between 0.8 -- 8 $R_\odot$. 
Furthermore, using an extended sample of $\sim$3900 {\it{Kepler}} mission stars, of which 
$\sim$300  stars are dwarfs and subgiants and $\sim$3600  are first-ascent giants ranging between 0.8 $R_\odot$ and 30 $R_\odot$, \citet{2019Zinn} reports  a 2 per cent agreement between radii based on {\it{{\it{Gaia}}}} parallax to radii based on asteroseismic scaling relations, further illustrating the accuracy and precision of scaling relations.

{\it{{\it{Gaia}}}} data allows stellar luminosities to be obtained based on their precise parallax measurements, -- providing an additional constraint whose impact in the model grid search needs to be examined.
The main aim of this article is two folds: we carry out a uniform characterisation of an ensemble of main-sequence benchmark stars and (i) assess the relevance of a stellar luminosity (i.e.\ derived based on parallax measurement) towards the determination of stellar parameters, mainly radius and mass; and
(ii) we determine the contribution of a model-independent radius measurement towards the estimation of a precise and accurate stellar mass. Finally, we examine the  precision needed for an interferometric radius (model-independent radius) to have an influential impact towards the determination of stellar mass in the optimisation process.
%
%
%
\section{Stellar model grid and benchmark stars}
\label{stellar}
\subsection{Stellar model grid}
\label{stellar_grid}
We employed the one dimensional (1D) stellar code  MESA (Modules for Experiments in Stellar Astrophysics \citealt{Pax1,Pax2,pax3,pax4,2019Paxton}) version r12778 for the computation of stellar models. The computed stellar model grid is made up of main-sequence stellar evolution tracks running from the zero-age main-sequence (ZAMS) phase to the end of the main-sequence phase, 
spanning the parameter space in mass, $M$, [0.7 -- 1.6] M$_\odot$ in steps of 0.05 M$_\odot$, initial metallicity, [Fe/H], [-0.50 -- 0.50] dex in steps of 0.1, and the helium enrichment ratios\footnote{We note that the range of helium enrichment ratio used also includes the Solar value of 1.23 deduced from our Solar calibration procedures.}, $\Delta Y/\Delta Z$, from [0.4 -- 2.4].  The initial helium mass fractions, $Y_{\rm{i}}$ is determined following the expression
\begin{equation}
   Y_{\rm{i}} = \left(\frac{\Delta Y}{\Delta Z} \right)Z_{\rm i} + Y_{0}~~,
   \label{helium}
\end{equation}
where $Z_{\rm{i}}$ is the initial metal mass fraction and $Y_{0}$ is the primordial
big bang nucleosynthesis helium mass fraction value taken as 0.2484 \citep{CYBURT2003227}.
For each stellar model, we calculated the corresponding adiabatic theoretical oscillation frequencies for spherical degrees $l$ = 0, 1, 2, and 3, using the GYRE oscillation code \citep{2013Townsend}.

The global input physics of our stellar grid includes  NACRE (Nuclear Astrophysics Compilation of REaction rates; \citealt{ANGULO19993}) reaction rates with specific rates for 
$^{14}\rm N(p,\gamma)^{15}\rm O$ 
described by \citet{Imbriani}. In addition, the grid used the 2005 updated version of the OPAL equation of state \citep{Rogers}. At high temperatures OPAL tables \citep{Iglesias} were used to cater for opacities, while tables from \citet{Ferguson} were used for lower temperatures. Furthermore, our grid takes into account solar chemical mixtures from \citet{2009Asplund}. The surface boundary of generated stellar models was described using the Krishna-Swamy atmosphere \citep{1966Krishna}. We also follow the prescription of \citet{cox1968} in the treatment of convection, and employ a solar calibrated  mixing length parameter, $\alpha_{\rm{MLT}}$ of 1.71. We note that atomic diffusion (gravitation settling component only; \citealt{1994Thoul}) was considered in stellar models which do not demonstrate over-depletion at the stellar surface (i.e.\ models $\lesssim$ 1.2 M$_\odot$). Lastly, for the stellar models with convective cores, we included convective core overshoot using the exponential decay with a diffusion coefficient described by \citet{2000Herwig}
\begin{equation}
    D_{\rm{ov}} = D_{\rm{0}}{\rm{exp}}\left(-\frac{z}{fH_{p}}     \right)~~,
    \label{ov}
\end{equation}
where $f$ is the overshoot parameter and restricted to 0.01, $H_{\rm{p}}$ is the pressure scale height, $D_{\rm{0}}$ is the diffusion coefficient of the unstable convective region, and $z$ is the distance from the boundary of the convective region.
\subsection{Benchmark stars}
\label{benchmark}
Our selected stellar sample is restricted to only main-sequence stars within the mass range [0.7 -- 1.6] M$_\odot$ 
and either having a model-independent radii and/or mass available. 
Table~\ref{table_stars1} shows the list of our benchmark stars, their corresponding adopted nomenclature, spectroscopic constraints (metallicities, [Fe/H], and effective temperatures, $T_{\rm{eff}}$), and interferometric radii. 
We also include the Sun-as-a-star among the benchmark stars. This is because the Sun is one of the best characterised stars which displays a rich spectrum of 
excited oscillation modes with extremely high signal-to-noise ratios.
The adopted solar frequencies are from \citet{2017Lund},
whose quality is similar to that of {\it{Kepler}} space mission. For the purpose of this work, this is relevant because it allows us to treat the Sun as a ``{\it{Kepler}} observed star'', hence assessing the precision and accuracy of the inferred parameters in a similar manner as for the other selected benchmark stars in Table~\ref{table_stars1}.  
\begin{table*}
\centering 
\caption{Selected benchmark stars, their commonly used  nomenclature, corresponding spectroscopic constraints (metallicities, [Fe/H], and effective temperatures, $T_{\rm{eff}}$) and interferometric radii, $R$. The second column represents the {\it{Kepler}} Input Catalogue (KIC) name while the third column indicates that adopted in the plots in this article.}
\begin{tabular}{cccccc}        
\hline 
\rowcolor{white}
Target star &  KIC & Used ID      & [Fe/H] (dex)   &	$T_{\rm{eff}}$ (K) & $R$ (R$_\odot$) 	 \\
\hline\hline
16~Cyg~A    & 12069424 & Cyg A       & 0.10 $\pm$ 0.03$^b$  & 5825 $\pm$ 50$^b$      & 1.220 $\pm$ 0.020$^a$\\
\rowcolor{gray!25}
16~Cyg~B    & 12069449 & Cyg B       & 0.05 $\pm$ 0.02$^b$  & 5750 $\pm$ 50$^b$ & 1.120 $\pm$ 0.020$^a$ \\
Doris      & 8006161    & Doris     & 0.34 $\pm$ 0.10$^c$  & 5466 $\pm$ 77$^c$   & 0.952 $\pm$ 0.021$^d$ \\
\rowcolor{gray!25}
Perky     & 6106415   & Perky       & -0.04 $\pm$ 0.10$^c$  & 6037 $\pm$ 77$^c$ & 1.289 $\pm$ 0.037$^d$\\
Saxo2    & 6225718    & Saxo2       &  -0.07 $\pm$ 0.10$^c$ & 6313 $\pm$ 76$^c$  & 1.306 $\pm$ 0.047$^d$                \\
\rowcolor{gray!25}
$\alpha$~Centauri~A & - & Cen A & 0.23 $\pm$ 0.05$^e$  & 5832 $\pm$ 62$^e$ & 1.2234 $\pm$ 0.0053$^e$ \\
$\alpha$~Centauri~B & - & Cen B & 0.23 $\pm$ 0.05$^e$  & 5795 $\pm$ 19$^e$ & 0.8632 $\pm$ 0.004$^e$ \\
Sun          & -    & Sun   & 0.00 $\pm$ 0.05  & 5777 $\pm$ 50$^g$ & 1.000 $\pm$ 0.0004$^f$ \\
\rowcolor{gray!25}
\hline                                   
\end{tabular}
\label{table_stars1}\\
$^{a}$\citet{2013White},$^b$\citet{2009Ram}, $^c$\citet{Buchhave_2015}, $^d$\citet{2012Hub}, {\bf{$^e$\citet{2018MNsamba}}}, $^f$\citet{Meftah2018},
$^g$\citet{2016Andrej}
\end{table*}
We now highlight the available essential stellar properties of each of our stars:
\begin{itemize}
    \item[---] 16~Cyg~A (HD 186408, HR 7503, KIC 12069424 ) and B (HD 186427, HR 7504, KIC 12069449) are binary star systems whose angular diameters have been measured by \citet{2013White}, based on observations using long-baseline optical interferometry with the Precision Astronomical Visible Observations (PAVO; \citealt{2008reland}) beam combiner  at the High Angular Resolution Astronomy (CHARA; \citealt{2005ten}) Array. \citet{2013White} found the angular diameters of 16~Cyg~A and B to be 0.539 $\pm$ 0.007 mas and 0.490 $\pm$ 0.006 mas, respectively. In addition, through the combination of interferometric diameters with {\it{Hipparcos}} parallax measurements, they measured a linear radius of 1.22 $\pm$ 0.02 R$_\odot$ and 1.12 $\pm$ 0.02 R$_\odot$ for 16~Cyg~A and B, respectively. Unfortunately, 16~Cyg~A and B  have long orbital periods estimated to be over 18,000 yr \citep{1999Hauser}, and therefore no dynamical masses are available. It is also worth noting that 16~Cyg~B hosts a Jovian-mass companion planet whose mass is estimated to be 1.5 MJ \citep{1997Cochran}.

    16~Cyg~A and B are well-studied solar analogues following the continuous {\it{Kepler}} space observation for about 2.5 yrs, providing high-quality seismic data, allowing for the extraction of over 48 oscillation frequency modes \citep{2017Lund}. This lead to a series of detailed seismic studies ranging from those involving testing asteroseismic tools which employ different optimisation techniques (such as machine leaning, Bayesian techniques; \citealt{2012Metcalfe,2020Farnir,2017Aguirre,2017Bellinger}), analysis of acoustic glitch signatures aimed at constraining the surface helium abundances (e.g. \citealt{Verma_2014}), to constraining core properties such as core hydrogen abundances \citep{2022Nsamba}.

    \item[---] $\alpha$~Centauri system is the brightest, closest triple star system to our Sun. It consists of  Proxima Centauri (HIP 70890) and a binary component made of solar-type stars, i.e.\ $\alpha$~Centauri~A (HD 128620, HR 5459) and B (HD 128621, HR 5460). This binary component provides a unique opportunity for testing and improving our understanding of stellar interior physics. This is attributed to the wealth of available highly precise observations, namely, interferometric radii ($R_{\rm A}$ = 1.2234 $\pm$ 0.0053 R$_\odot$ and $R_{\rm B}$ = 0.8632 $\pm$ 0.004 R$_\odot$; \citealt{2017Kervella}), dynamical masses ($M_{\rm A}$ = 1.1055 $\pm$ 0.004 M$_\odot$ and $M_{\rm B}$ = 0.9373 $\pm$ 0.003 M$_\odot$; \citealt{2017Kervella}), precise parallax measurements (747.1$ \pm$ 1.2; \citealt{Staffan}, 743 $\pm$ 1.3; \citealt{2016Pourbaix}, 747.12 $\pm$ 0.61; \citealt{2016vella}), and asteroseismic data from a handful of ground-based surveys (\citealt{2002Bouchy,Bedding_2004,Kjeldsen_2005,2005Bedding,10Bazot,2010Meulenaer}). Table~\ref{table_stars1} also shows the metallicity and effective temperatures adopted for $\alpha$ Centauri A and B. 

    $\alpha$~Centauri binary system has been used as a test-bed for stellar model physics because of the precisely available observational constraints \citep{2004Eggenberger,2005Miglio,2007Y,2018Joyce}. In fact, enormous efforts have been dedicated towards ascertaining the nature of the core of one of the binary components, $\alpha$~Centauri~A. This is because its dynamical mass (~1.1 M$_\odot$) lies in the mass region where stars are expected to develop a convective core while on the main-sequence, i.e.\ 1.1 -- 1.2 M$_\odot$ (see details in \citealt{10Bazot,2016Bazot,2018MNsamba,10Nsamba,2021Salmon}). The disagreements in the predictions of the convective/radiative nature of the core of $\alpha$~Centauri~A is expected to be resolved when more precise oscillation frequencies are made available.

    \item[---] KIC 8006161 (HD 173701) also famously known as Doris, is a metal-rich seismic solar analog with both spectroscopic ground-based observations (\citealt{2018AKaroff}) and up-to 2.5 yrs of continuous {\it{Kepler}} observations \citep{2017Lund}. The angular diameter measurements of Doris have been measured by \citet{2012Hub}, yielding an interferometric radius of 0.952 $\pm$ 0.021 R$_\odot$. Doris is also reported to have a high metallicity of $\rm{[Fe/H]} = 0.34 \pm 0.1$~dex and an effective temperature of 5488 $\pm$ 77 K \citep{Buchhave_2015}. These sets of constraints and characteristics of Doris make it an interesting benchmark solar-like main-sequence star. Due to its high metallicity, various studies have not only focused on deriving the fundamental stellar properties of Doris (e.g. \citealt{2018ApJBasu}), but have also been directed towards examining its magnetic activity cycle (\citealt{2017A&AKiefer,2018Santos,Kiefer2020,2023Santos}). The latter is because of the variations in opacity and convection zone depth attributed to its high metallicity.

    \item[---] KIC 6106415 (HD 177153) and KIC 6225718 (HD 187637) are commonly known as Perky and Saxo2, respectively. These two stars are part of the {\it{Kepler}} Legacy sample stars with high quality sesimic data \citep{2017Lund,2017Aguirre}. \citet{2012Mathur} analysed a group of 22 stars (including KIC 6106415) using the Asteroseismic Modelling Portal (AMP: \citealt{2003JCoPh.185..176M,calfe2023}). Interestingly, they reported the initial helium abundance of KIC 6106415 to be consistent with the primordial helium abundance of 0.246, while \citet{2019Verma} determined the envelope helium abundance of KIC 6106415 through the analysis of
    a glitch signature arising from the ionization of helium and found it to span the range [0.201 -- 0.222]. This disagreement in the initial helium abundance has a direct impact on the inferred stellar mass as illustrated in \citet{2021Nsamba}. The interferometric radii measurements of KIC 6106415 and KIC 6225718 are available from \citet{2012MHuber}, i.e.\ 1.289 $\pm$ 0.037 R$_\odot$ and 1.306 $\pm$ 0.047 R$_\odot$, respectively. KIC 6106415 has a metallicity of -0.04 $\pm$ 0.1 dex and effective temperature of 6037 $\pm$ 77 K \citep{Buchhave_2015}. KIC 6225718 is also reported to have a metallicity of -0.07 $\pm$ 0.1 dex and effective temperature of 6313 $\pm$ 76 K \citep{Buchhave_2015}.    
\end{itemize}

\subsection{{\it{{\it{Gaia}}}} parallax-based luminosities and radii}
\label{parallax_lum}
Table~\ref{parallax} shows the {\it{\bf{{\it{Gaia}}}}} parallax measurements for our stellar sample. Among the benchmark stars, $\alpha$~Centauri~A and B don't have any {\it{\bf{{\it{Gaia}}}}} parallax values available. This is attributed to the high brightness and binary nature of the  $\alpha$~Centauri~A and B system which limits the absolute accuracy of {\it{\bf{{\it{Gaia}}}}} custom data reduction processes \citep{2021Akeson}. Therefore, we adopt the available parallax measurements of $\alpha$~Centauri~A and B from {\it{Hipparcos}} data \citep{2007Leeuwen}.

\begin{table*}
\caption{{
Parallax-based luminosities for our sample stars. The eighth column shows the radii values deduced in this work using equation~{\ref{radi}}}
}
\centering
\begin{tabular}{cccccccc}
\hline
\rowcolor{white}
Target Star & $\pi_{\rm{{\it{Gaia}}}}$ (mas) & $\pi_{\rm{{\it{Hipparcos}}}}$ (mas) & $V_{\rm{mag}}$ & $E(B-V)$  & $L_{\rm{{\it{Gaia}}}}$ (L$_{\odot}$) & $L_{\rm{{\it{Hipparcos}}}}$ (L$_{\odot}$) & $R$ (R$_\odot$)\\
\hline
16~Cyg~A    &   47.32 $\pm$ 0.02   &   47.44 $\pm$ 0.27   &   5.99   &   0.001 $\pm$ 0.015   &   1.50 $\pm$ 0.06   &   1.49 $\pm$ 0.07 & 1.225 $\pm$ 0.014\\
\rowcolor{gray!25}
16~Cyg~B    &   47.33 $\pm$ 0.02   &   47.14 $\pm$ 0.27   &   6.25   &   0.001 $\pm$ 0.015   &   1.19 $\pm$ 0.05   &   1.20 $\pm$ 0.05 & 1.113 $\pm$ 0.014\\
Doris    &   36.99 $\pm$ 0.02   &   37.47 $\pm$ 0.49   &   7.54   &   0.002 $\pm$ 0.015   &   0.63 $\pm$ 0.03   &   0.61 $\pm$ 0.03 & 0.965 $\pm$ 0.017 \\
\rowcolor{gray!25}
Perky    &   24.16 $\pm$ 0.01   &   24.11 $\pm$ 0.44   &   7.20   &   0.004 $\pm$ 0.015   &   1.85 $\pm$ 0.08   &   1.86 $\pm$ 0.10 & 1.286 $\pm$ 0.0267\\
Saxo2    &   19.01 $\pm$ 0.02   &   19.03 $\pm$ 0.46   &   7.53   &   0.004 $\pm$ 0.015   &   2.14 $\pm$ 0.09   &   2.14 $\pm$ 0.14 &  1.307 $\pm$ 0.034\\
\rowcolor{gray!25}

$\alpha$~Centauri~A     &   -       &   742.12 $\pm$ 1.40   &   -0.01   &   0.000 $\pm$ 0.014   &   -   &   1.53 $\pm$ 0.06 & - \\
$\alpha$~Centauri~B    &   -       &   742.12 $\pm$ 1.40   &   1.35  &   0.000 $\pm$ 0.014   &   -   &   0.50 $\pm$ 0.02 & - \\
\hline
\end{tabular}
\label{parallax}
\end{table*}

\begin{figure}
    \includegraphics[width=\columnwidth]{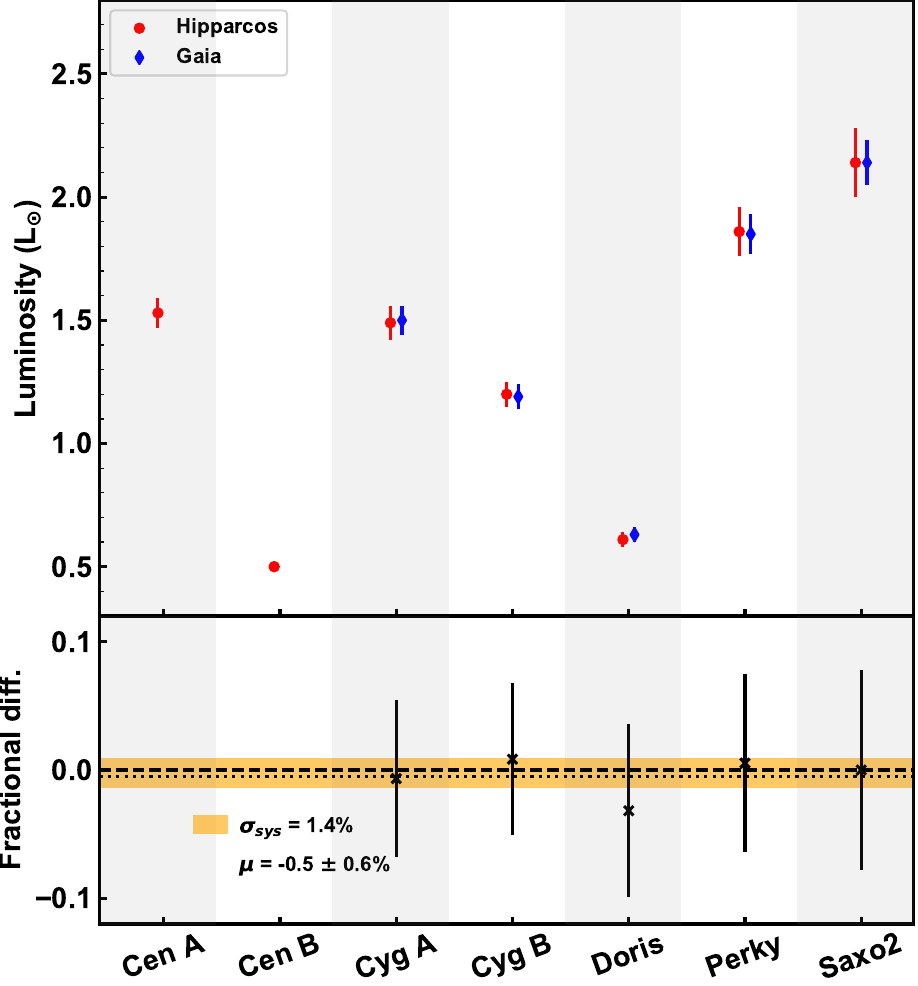}
      \caption{
      Top panel shows the comparison of the absolute parallax-based luminosities values and their corresponding uncertainties from {\it{Hipparcos}} (red circles) and {\it{Gaia}} (blue diamonds). The bottom panel shows the fractional difference in luminosity relative to the {\it{Gaia}}-based luminosity value, with a scatter of 1.4 per cent (orange color), and an offset of $-0.5 \pm 0.6$ per cent (dotted black line). No {\it{Gaia}} parallaxes are available for $\alpha$ Centauri A and B.
     }
    \label{luminosity_comparison}
\end{figure}
The parallax-based luminosities ($L$) for our benchmark stars were calculated using the expression \citep{Pijpers}
\begin{equation}
    {\rm{log}}\left(\frac{L}{L_{\odot}}   \right) = 4 + 0.4M_{\rm{bol,\odot}} - 2{\rm{log}}\pi{\rm[mas]} - 0.4(V - A_{\rm v} + {\rm{BC_v}})~,
    \label{luminosity_eqn}
\end{equation}
where $M_{\rm{bol,\odot}}$ is the Solar bolometric magnitude approximated to 4.73 based on \citet{2010Torres},  $\pi_{\rm{{\it{Gaia}}}}$ and $\pi_{\rm{{\it{Hipparcos}}}}$ are parallaxes obtained from {\it{\bf{{\it{Gaia}}}}} DR3 data release\footnote{https://gea.esac.esa.int/archive/} and {\it{Hipparcos}} data\footnote{http://vizier.cds.unistra.fr/viz-bin/VizieR?-source=V/137D/XHIP}, respectively. 
For comparison purposes, we also determine the luminosity values for our target stars based on their corresponding {\it{Hipparcos}} parallax measurements (see Table~\ref{parallax} and Figure~\ref{luminosity_comparison}). 
$V$ is the visual magnitude obtained from the same {\it{Hipparcos}} database. The bolometric corrections, $\rm{BC_v}$, were calculated from polynomials expressed as functions of stellar effective temperatures as suggested by \citet{1996Flower} and corrected by \citet{2010Torres}. 
The stellar extinction in the V band, $A_{\rm v}$, was determined using the expression 
\begin{equation}
    A_{\rm v} = R \times E({B-V}) ~,
    \label{extinction}
\end{equation}
where $R$ is the reddening constant of 3.1 adopted based on \citet{Fitzpatrick_1999} and $E({B-V})$ is the colour excess, obtained using STILISM\footnote{https://stilism.obspm.fr/} (STructuring by Inversion the Local Intersteller Medium: \citealt{2014Lallement,2017Capitanio}).
%
%

Figure~\ref{luminosity_comparison} shows a comparison of the derived luminosities based on {\it{\bf{{\it{Gaia}}}}} and {\it{Hipparcos}} parallaxes, 
with the bottom panel showing their fractional differences with a scatter of $\sim$1.4 per cent and an offset of $-0.5 \pm 0.6$ per cent. From Table~\ref{parallax}, it is evident that the {\it{\bf{{\it{Gaia}}}}} parallax measurements are more precise compared to the {\it{Hipparcos}} parallaxes, 
however, this barely translates into somewhat more precise {\it{\bf{{\it{Gaia}}}}}-based luminosities compared to {\it{Hipparcos}}-based luminosities. This is because errors in $A_{\rm v}$ (determined based on equation~\ref{extinction}) dominate the parallax errors during error propagation in equation~(\ref{luminosity_eqn}).
%

Taking into account the {\it{Gaia}} parallaxes (shown in Table~\ref{parallax}), and the angular diameter measurements of our benchmark stars (available from \citealt{2012Hub} and \citealt{2013White}), we compute the interferometric radius measurements using the expression
\begin{equation}
    R = \frac{1}{2} \theta_{\rm LD} D~,
    \label{radi}
\end{equation}
where $\theta_{\rm LD}$ is the angular diameter and $D$ is the distance to the star, which is calculated directly from the parallax. The corresponding radius uncertainties were determined using the expression
\begin{equation}
    \sigma (R) = R \times \sqrt{\left(\frac{\sigma(\theta_{\rm{LD}})}{\theta_{\rm{LD}}}  \right)^2 + \left(\frac{\sigma (D)}{D}       \right)^2
    }  ~~,          
\end{equation}
where $\sigma (\theta_{\rm LD})$ is the uncertainty on the angular diameter and $\sigma (D)$ is the uncertainty on the distance to the star.  

\subsection{Optimisation process}
\label{optimisation}
In order to infer stellar parameters (mass and radius), we make use of the AIMS (Asteroseismic Inference on a Massive Scale; \citealt{2019Rendle})\footnote{https://lesia.obspm.fr/perso/daniel-reese/spaceinn/aims/version2.0/} code, which relies on a pre-computed grid of models, and fits a specified set of atmospheric and seismic constraints using an MCMC (Markov Chain Monter Carlo; \citealt{Gilks,Dani}) approach. 
The advantage of the AIMS code over various seismic optimisation codes
is that it allows for interpolation within a pre-computed grid of models through tessellation/triangulation of the parameter space, thus enabling the identification of stellar evolutionary tracks located within the parameter space. 
Details are available in the AIMS documentation\footnote{https://sasp.gitlab.io/aims/} and \citet{2019Rendle}.

In a nutshell, AIMS allows for a generation of a subset of models which are representative of the specified constraints from which posterior distributions of the different model properties 
(such as mean density, radius, mass, age, among others)
and their corresponding standard deviations, percentile ranges can be determined. The $\chi^2$ is handled following the general expression
\begin{equation}
    \chi^2 = \sum_{l=1}^{N} \left( \frac{A_i - B_i}{\sigma_i} \right)^2~,
    \label{general_chi_square}
\end{equation}
where $A_i$, $B_i$, and $\sigma_i$ are the observed parameters, model parameters, and corresponding associated observed uncertainties, respectively. The $\chi^2$ in equation~(\ref{general_chi_square}) is a composition of the seismic and atmospheric components expressed as
\begin{equation}
        \chi^2 = \eta\left(\chi^2_{\rm{seismic}}\right) + \chi^2_{\rm{atmospheric}}~,
        \label{chi_total}
\end{equation}
where $\eta$ = $N_{a}/N_{\nu}$, the ratio of the number of atmospheric constraints, $N_{a}$, to seismic constraints, $N_{\nu}$. The inclusion of $\eta$ in equation~(\ref{chi_total}) facilities for the specification of equal weights, thus yielding the same weight for every observable in the likelihood function. It is important to note that depending on the approach adopted, the weights given to seismic and atmospheric constraints have an impact on the inferred stellar properties and their corresponding statistical uncertainties. Refer to \citet{2021Cunha} for an indepth exploration of the impact of applying different weights on the derived stellar parameters arising from employing various model selection processes.
The seismic $\chi^2$ component in equation~(\ref{chi_total}) makes use of the observed oscillation frequencies, $\nu_{\rm{obs}}$, their corresponding uncertainties, $\sigma (\nu)$, and the theoretical model oscillation frequencies, $\nu_{\rm{mod}}$, taking the form
\begin{equation}
    \chi^2_{\rm{seismic}} = \sum_{l=1}^{N} \left( \frac{\nu^{\rm{obs}} - \nu^{\rm{mod}}}{\sigma (\nu)} \right)^2~.
    \label{general chi_seismic}
\end{equation}
We note that the disparity between the observed oscillation frequencies and the model oscillation frequencies arising from the improper modeling of near-surface layers, also known as the ``surface effects'' \citep{1988Christensen,1988Dziembowski,1997Christensen-Dalsgaard}, was rectified using the two-term surface correction empirical formula suggested by \citet{2014Ball}. A detailed comparison of the performance of different surface correction routines is given in \citet{2018MNRANsamba}, \citet{2018Compton}, \citet{2020rgensen}, among others.
The atmospheric $\chi^2$ component takes into account the available atmospheric contraints (i.e.\ effective temperature, $T_{\rm{eff}}$, metallicity, [Fe/H], luminosity, $L$) expressed as
\begin{equation}
    \chi^2_{\rm{atmospheric}} = \chi^2_{T_{\rm{eff}}} + \chi^2_{\rm{[Fe/H]}} + \chi^2_{L} ~,
    \label{atmos_only}
\end{equation}
where $\chi^2_{T_{\rm{eff}}} = \left( \frac{T_{\rm{eff}}^{(\rm{obs})} - T_{\rm{eff}}^{(\rm{mod})}}{\sigma(T_{\rm{eff}})}  \right)^2$,
$\chi^2_{\rm{[Fe/H]}} = \left( \frac{T_{\rm{eff}}^{(\rm{obs})} - T_{\rm{eff}}^{(\rm{mod})}}{\sigma(T_{\rm{eff}})}  \right)^2$, and
$\chi^2_{L} = \left( \frac{{L}^{(\rm{obs})} - {L}^{(\rm{mod})}}{\sigma(L)}  \right)^2$~.
In cases where an interferometric radius is available, we assess its contribution towards the determination of a precise stellar mass 
by replacing the $\chi^2_L$ in equation~(\ref{atmos_only}) with 
 the radius $\chi^2$ which takes the form
$\chi^2_{R} = \left( \frac{{R}^{(\rm{obs})} - {R}^{(\rm{mod})}}{\sigma(R)}  \right)^2$~.
The superscripts ``obs'' and ``mod'' in the definitions of the respective $\chi^2$ terms of equation~(\ref{atmos_only}) correspond to the observed and model parameters.

The scatter (systematic uncertainties, $\sigma_{\rm{sys}}$) on the inferred stellar parameters were determined using the expression
\begin{equation}
    \sigma_{\rm{sys}} = \sqrt{\frac{1}{N} \left( \sum_i (x_i  -  \mu)^2    \right)}~~,
\end{equation}
where $N$ corresponds to the total number of quantities while $x_i$ denotes the fractional differences between quantities.
%
%
The mean (offset, $\mu$) and it's corresponding error ($\sigma_{\mu}$) are obtained using the expression
\begin{equation}
    \mu = \sum \frac{x_i}{N} 
\end{equation}
and 
\begin{equation}
    \sigma_\mu = \frac{\sigma_{\rm{sys}}}{\sqrt{N}}  ~~, 
\end{equation}
respectively. This aids in establishing the accuracy level of the inferred parameters.

\section{Results and discussions}
\label{results}
%
\subsection{Impact of a stellar luminosity on the inferred radius and mass}
\label{luminosity_results}
\begin{figure}
    \includegraphics[width=\columnwidth]{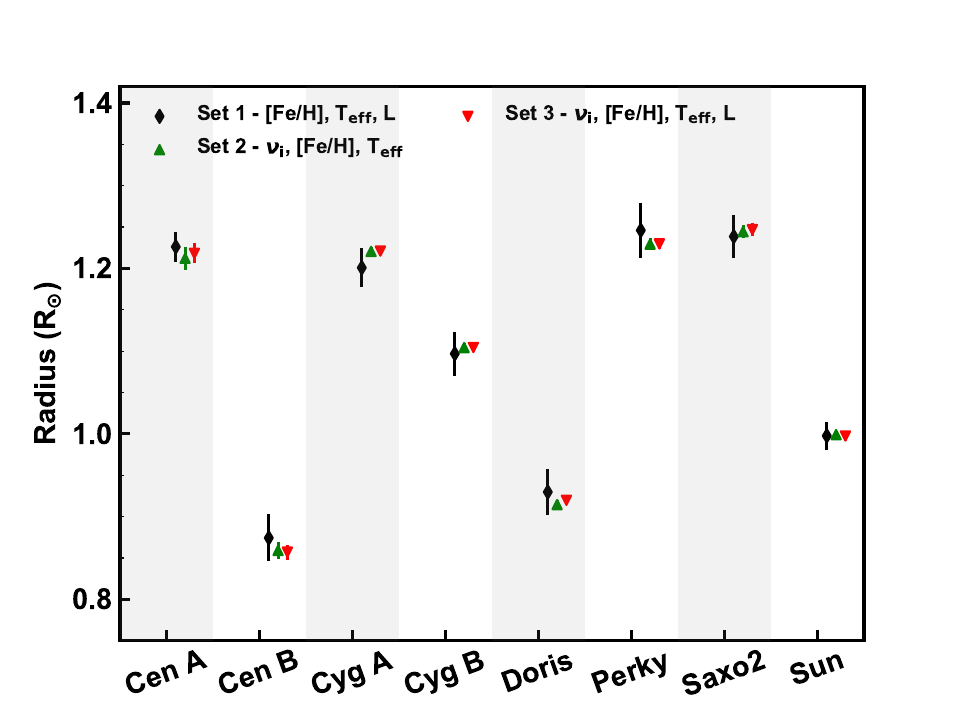}
    \includegraphics[width=\columnwidth]{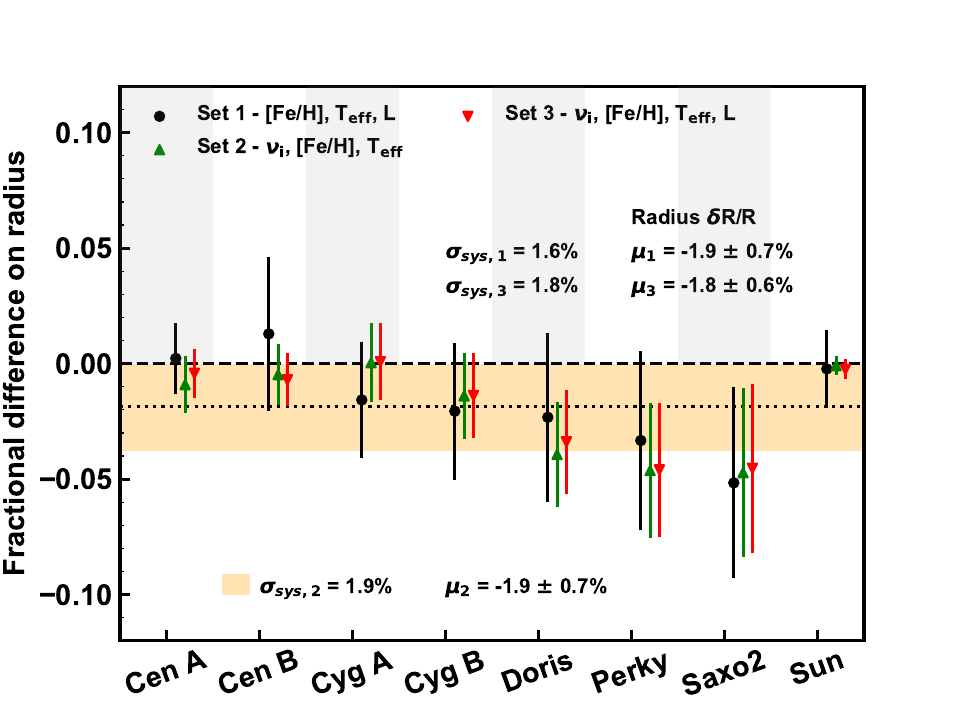}
      \caption{
      Top panel: comparison of the derived absolute radii and their associated uncertainties from different observable combinations. Bottom panel: fractional difference in radius relative to the interferometric radius. Orange color and dotted black line display the scatter ($\sigma_{sys, 2}$) and an offset ($\mu_2$) based on Set 2, respectively. For comparison purposes, the values for the scatter and offset based on Set 1 and Set 3 are also included.
      } 
    \label{Radius_L_Comp}
\end{figure}

\begin{figure}
    \includegraphics[width=\columnwidth]{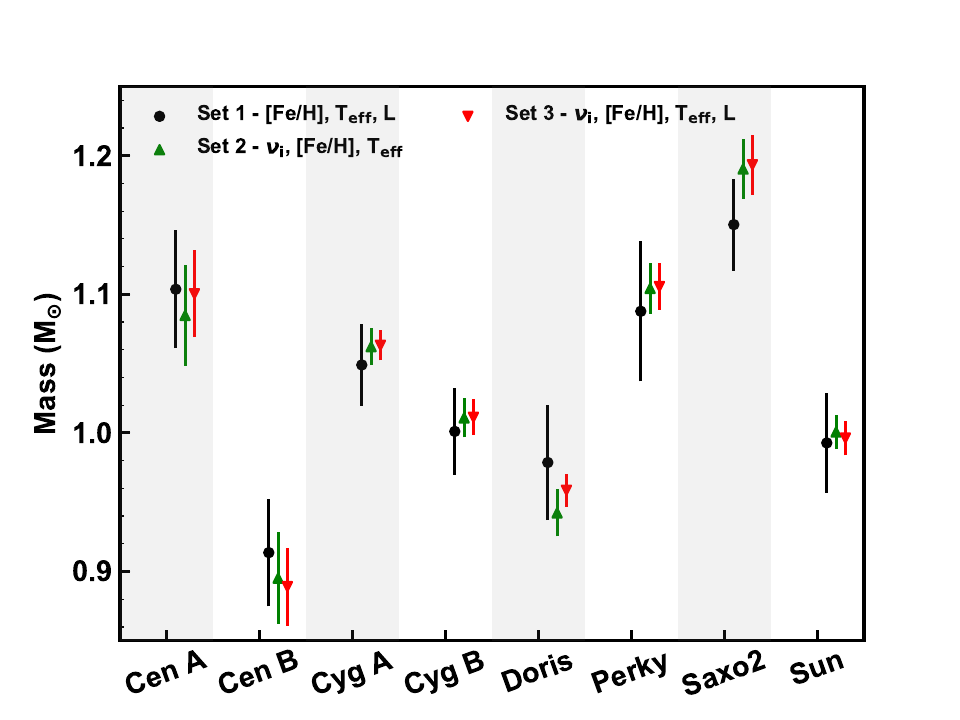}
 \includegraphics[width=\columnwidth]{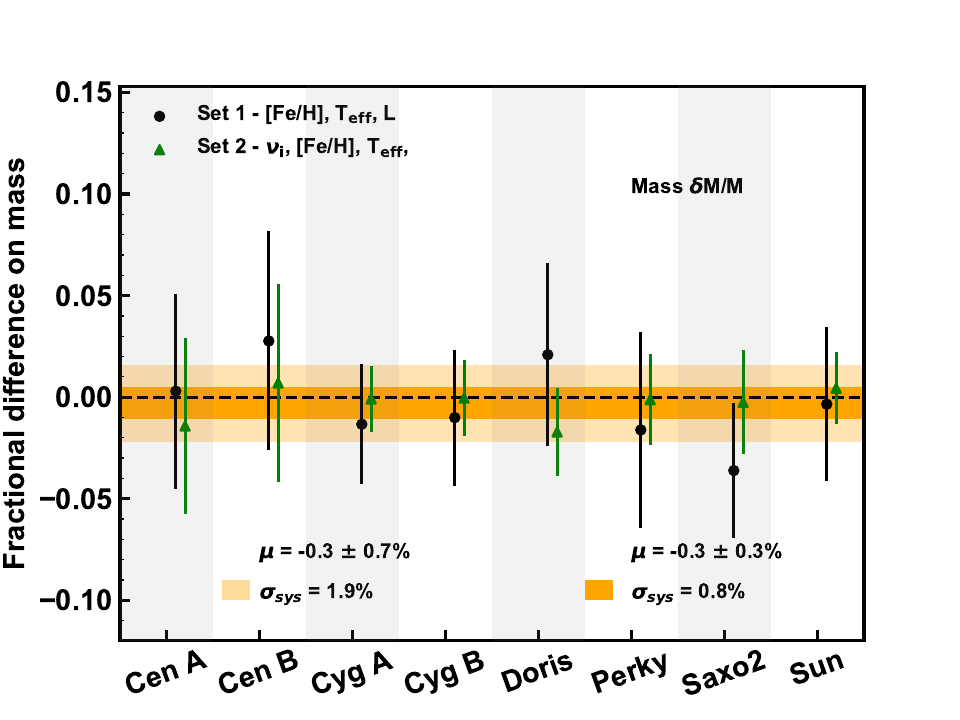}
      \caption{
      Top panel: comparison of the derived absolute masses and their associated uncertainties from different observable combinations. Bottom panel: fractional difference in mass derived using other Sets relative to Set~3. Light orange color shows the scatter (systematics) and corresponding offset (bias) when Set~1 is compared with Set~3, while the dark orange color compares Set~2 and Set~3.
      } 
    \label{Mass_Comp_L}
\end{figure}

\begin{figure*}
    \includegraphics[width=0.97\columnwidth]{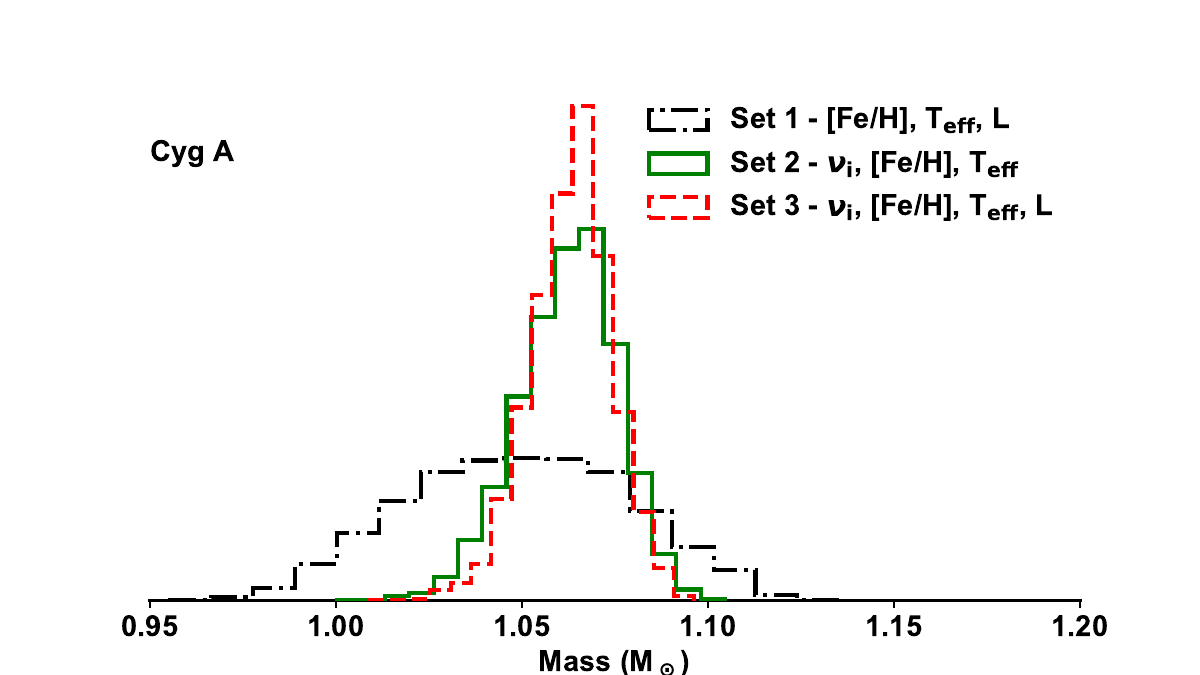}
    \includegraphics[width=0.97\columnwidth]{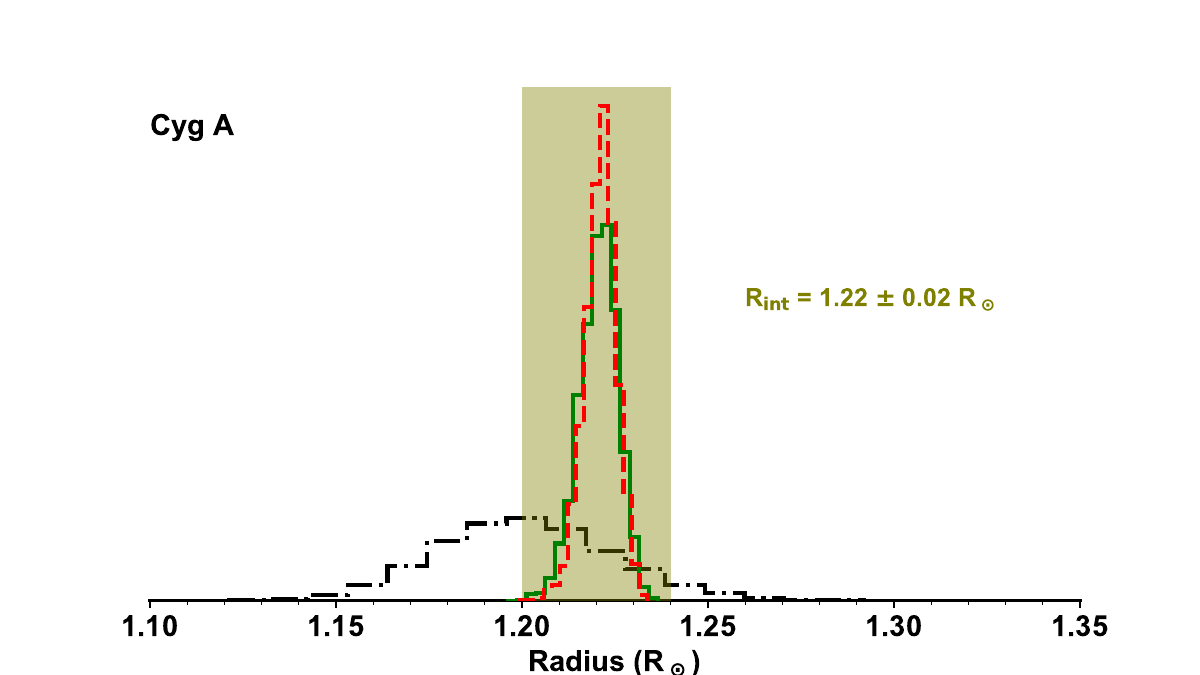}
    \includegraphics[width=0.97\columnwidth]{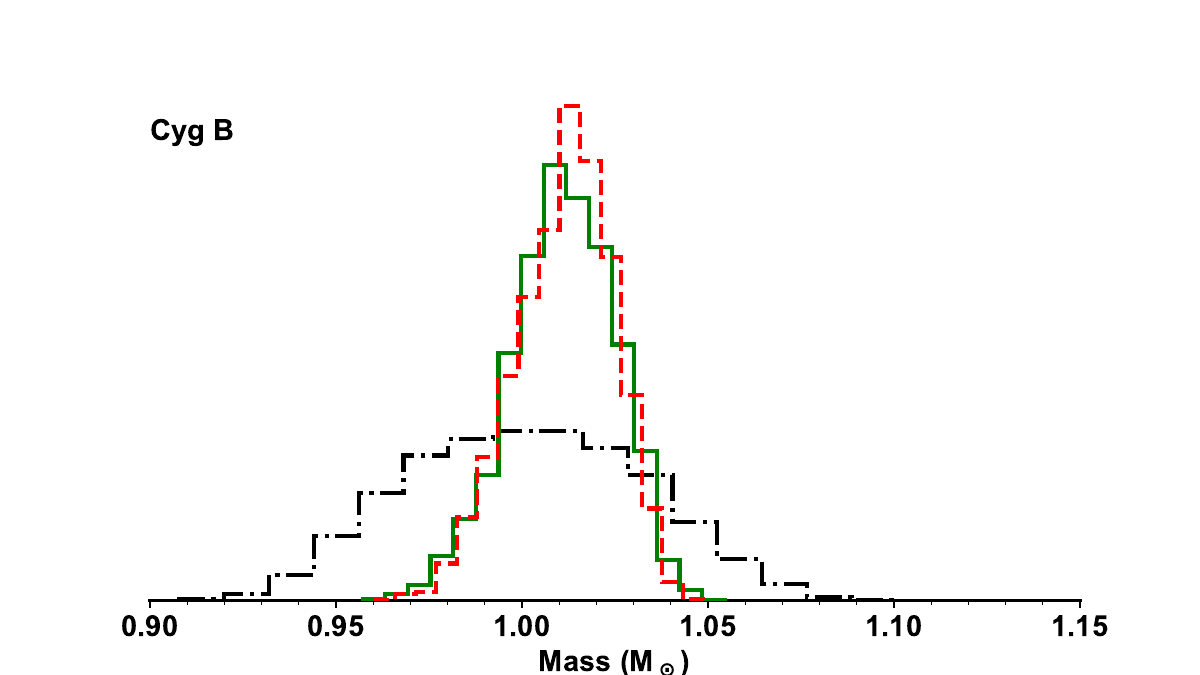}
    \includegraphics[width=0.97\columnwidth]{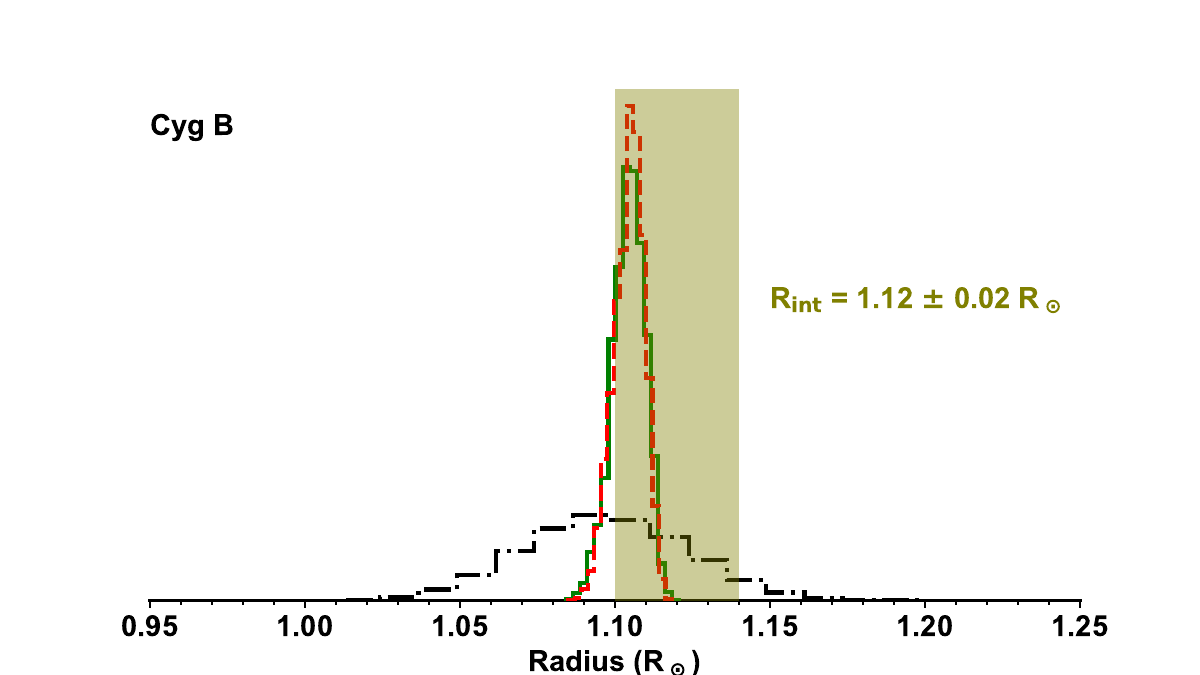}
    \includegraphics[width=0.97\columnwidth]{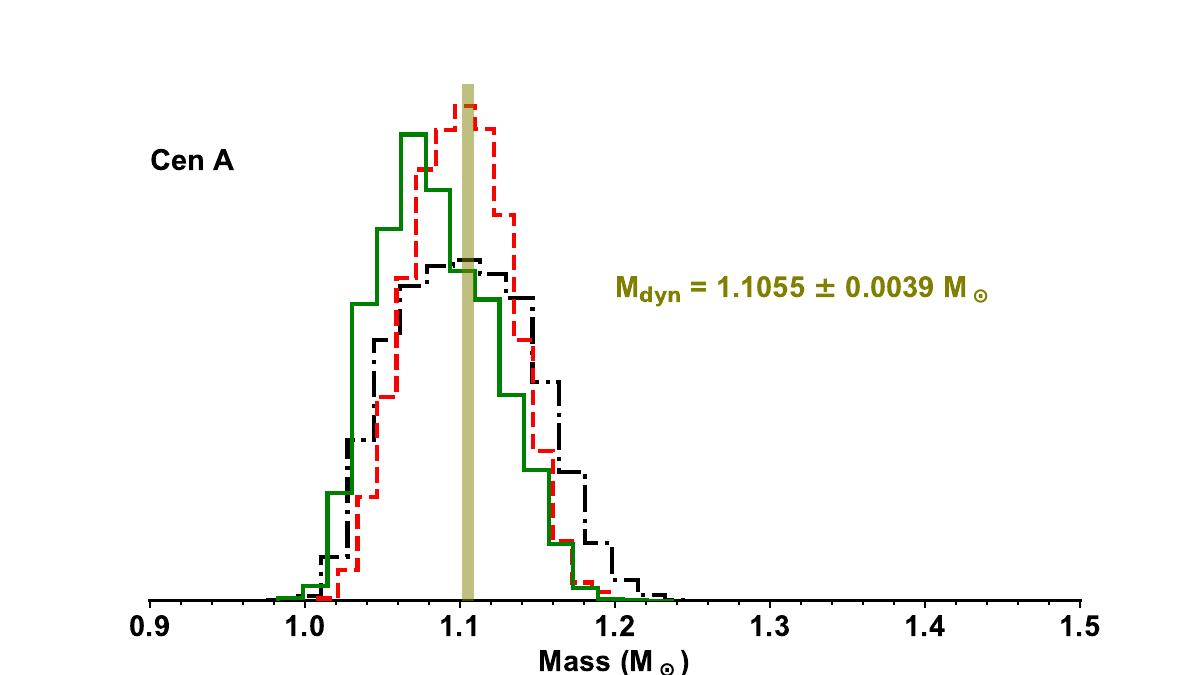}
    \includegraphics[width=0.97\columnwidth]{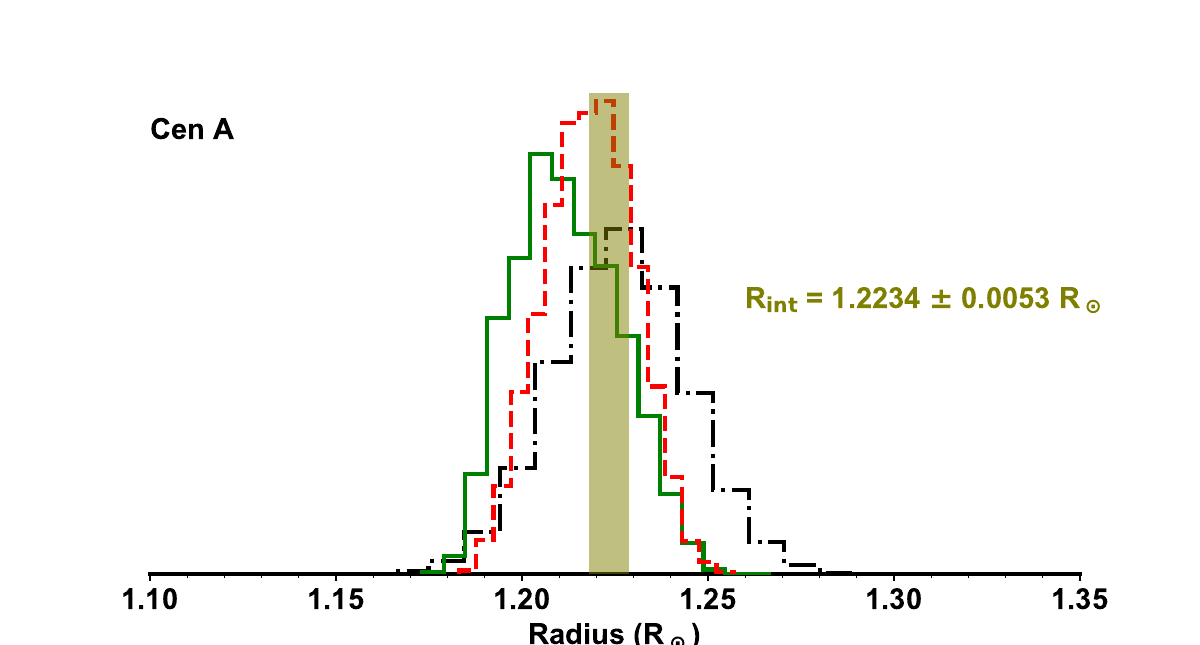}
    \includegraphics[width=0.97\columnwidth]{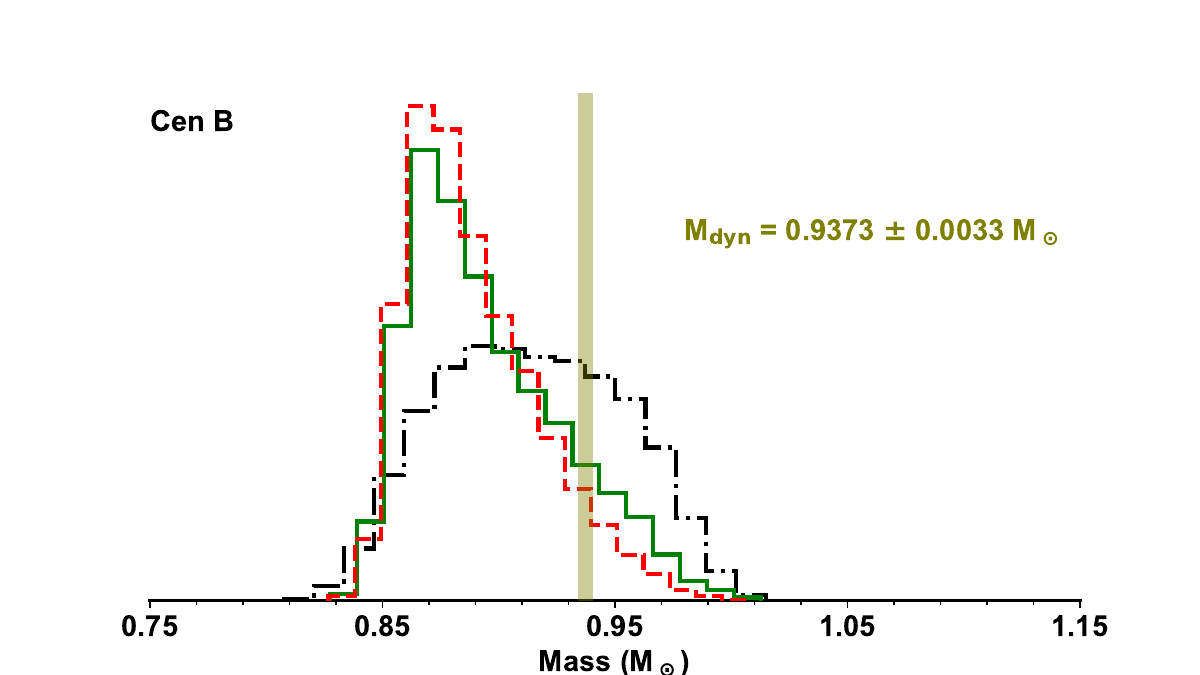}
    \includegraphics[width=0.97\columnwidth]{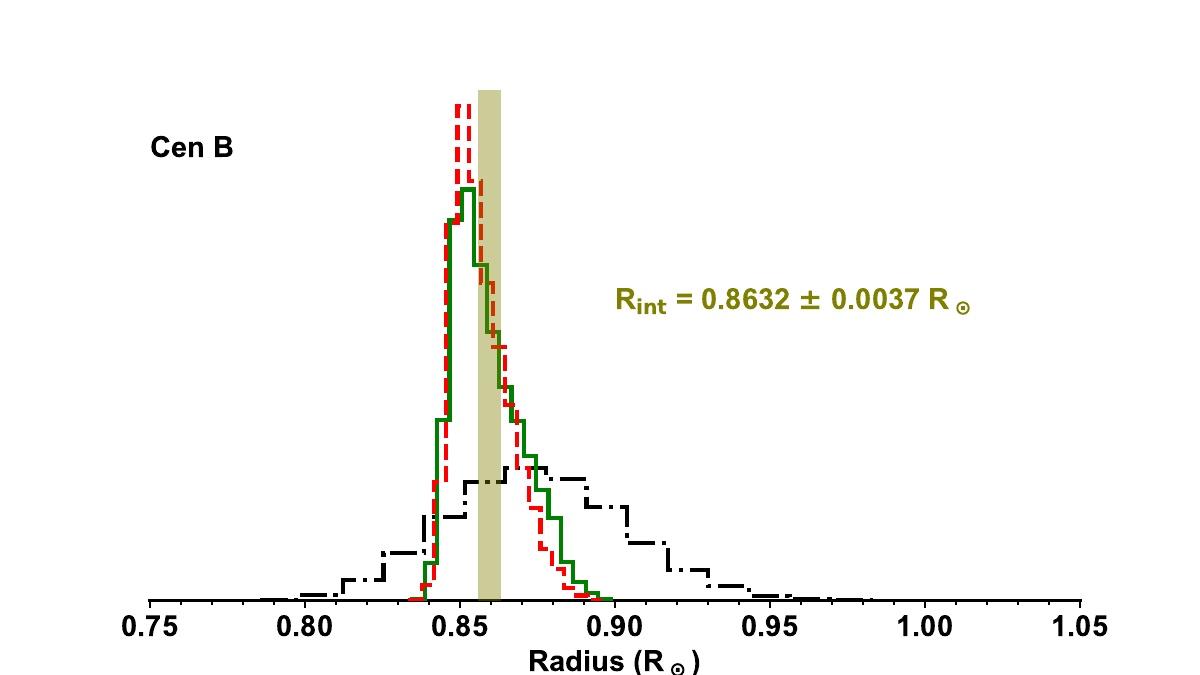}
    \includegraphics[width=0.97\columnwidth]{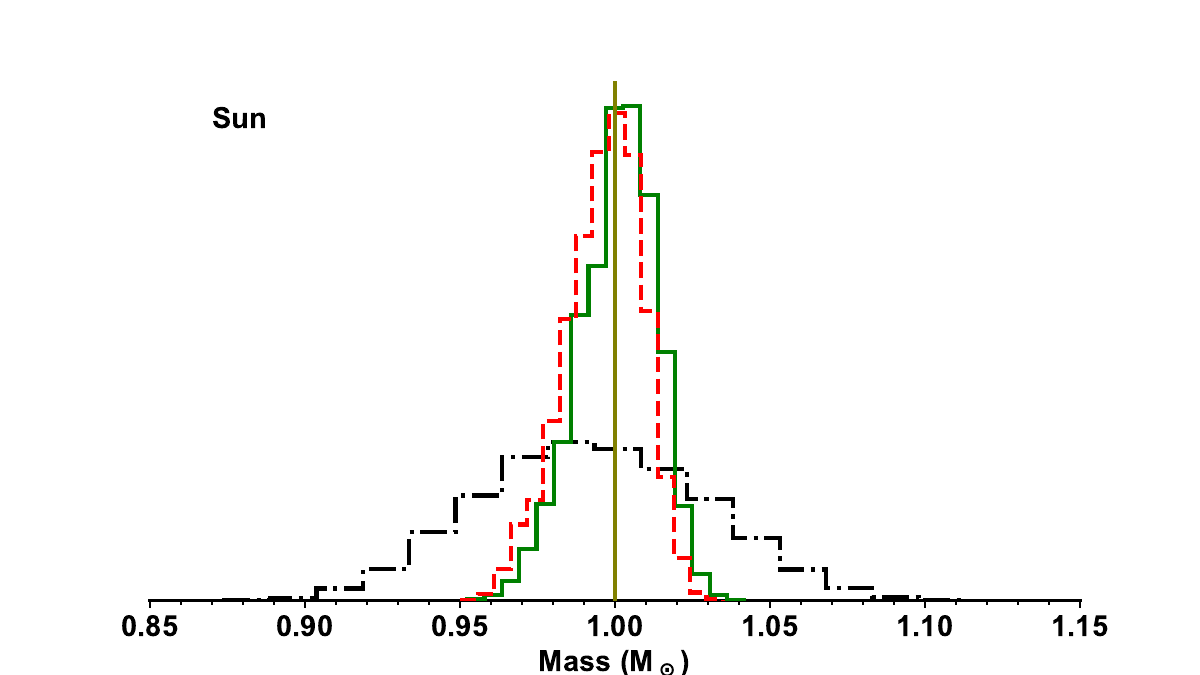}
    \includegraphics[width=0.97\columnwidth]{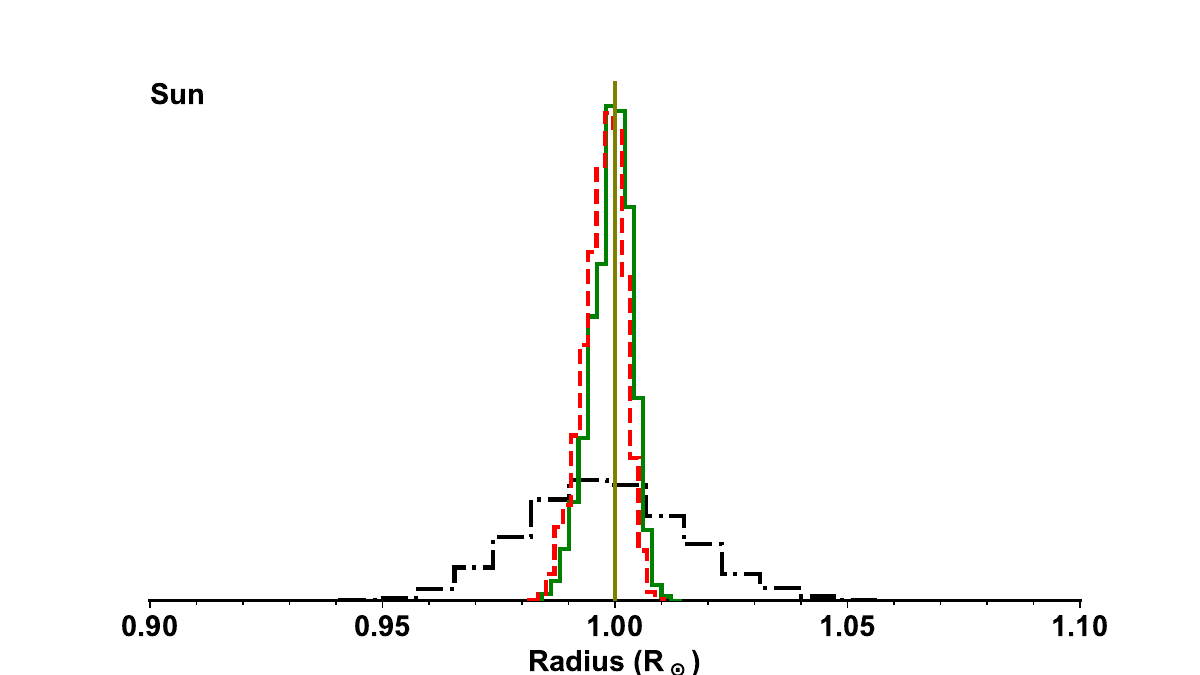}
      \caption{
      Normalised probability density distributions for mass (left panels) and radius (right panels). Color-code and line style shows results obtained using different observable combinations, i.e.\ Set~1 (black dash dotted line), Set~2 (green line), and Set~3 (red dashed line). The olive region given in some cases refers to independent values of radius and mass, if available.
       }     
    \label{pdf_radius_mass_compare}
\end{figure*}

%
%
Table~\ref{constraints} shows a combination of different atmospheric and seismic constraints used in establishing the relevance of a stellar luminosity towards the determination of stellar radius and mass. 
The top panel of Figure~\ref{Radius_L_Comp} and Table~\ref{radii_values} show that the inferred stellar radii  from Set~1 yields the least precise values compared to Set~2 and Set~3.
This is because constraints in Set~1 allow for a combination of an effective temperature and luminosity via the Stefan-Boltzmann's relation \citep{1884Boltzmann,2015Paul,2018Montambaux}, thus placing a restriction on the model radius selection process and consequently the mass. Set~2 and Set~3 include seismic data as an additional constraint, provided via individual oscillation frequencies in the optimisation process, putting strong restriction on the stellar mean density and radius.
Similar findings were reached in terms of precision of the inferred masses (see top panel of Figure~\ref{Mass_Comp_L}; black circle symbols and Table~\ref{masses_values}), except for $\alpha$~Centauri~A and B. 
This stems from the ground-based seismic data (for $\alpha$~Centauri~A and B) which is not as precise as the {\it{Kepler}} space based seismic data used for the remaining sample of our benchmark stars \citep{2003A&ACarrier,2010Meulenaer,2017Lund}. Furthermore, the mass and radius probability distributions become slightly more narrow as the constraints where changed from Sets 2 to 3 (see left and right panels of Figure~\ref{pdf_radius_mass_compare}). The combinations of observables in Sets 3 produce higher mass and radius probabilities followed by Set~2. This demonstrates that the addition of a parallax-based luminosity contributes vital information in finding the most probable stellar parameter.

\begin{table}
\centering 
\caption{
Different combinations of seismic and atmospheric constraints.
 }
\begin{tabular}{cc}        
\hline 
Sets         & observational combinations \\
\hline\hline
1            &  [Fe/H], $T_{\rm{eff}}$, $L$ \\
\rowcolor{gray!25}
2            &   $\nu_i$, [Fe/H], $T_{\rm{eff}}$ \\
3            &  $\nu_i$, [Fe/H], $T_{\rm{eff}}$,  $L$ \\

\hline                                   
\end{tabular}
\label{constraints}
\end{table}
The bottom panel of Figure~\ref{Radius_L_Comp} highlights that
the inferred stellar radii for majority of our stellar sample are underestimated compared to their interferometric radii values, yielding an offset of up-to $-1.9 \pm 0.7$ per cent and a scatter of $\sim$1.9 per cent. We note that this scatter relatively reduces for combinations involving luminosity as an additional constraint, i.e. $\sim$1.6 per cent and $\sim$1.8 percent for Set 1 and Set 2, respectively. 
Since no independent stellar masses are available for our binary stars, except for $\alpha$ Centauri A and B, we considered the inferred masses from Set~3 as a reference in the bottom panel of Figure~\ref{Mass_Comp_L}. Our findings show that when the masses from Set~1 (only atmospheric constraints) are compared with those from Set~2 (commonly used combination of seismic and classical constraints), and Set~2 compared with Set~3 (includes a stellar luminosity ), an offset of $-0.3$ per cent is attained, while the scatter is reduced from 1.9 per cent to 0.8 percent, respectively.


%
%
\begin{table*}
\centering 
\caption{
Inferred stellar radii from the different observable combinations. For the definition of included observables in each set, see Table~\ref{constraints}.
}
\begin{tabular}{ccccc}        
\hline 
Star              & Set~1  & Set~2 & Set~3 \\
\hline\hline
Cyg A             & 1.201 $\pm$ 0.023  & 1.221 $\pm$ 0.005  & 1.221 $\pm$ 0.004 \\
\rowcolor{gray!25}
Cyg B             & 1.097 $\pm$ 0.026  & 1.104 $\pm$ 0.005  & 1.105 $\pm$ 0.005 \\
Doris             & 0.929 $\pm$ 0.028  & 0.915 $\pm$ 0.006  & 0.920 $\pm$ 0.004 \\
\rowcolor{gray!25}
Perky             & 1.246 $\pm$ 0.033  & 1.229 $\pm$ 0.007 & 1.230 $\pm$ 0.007 \\
Saxo2             & 1.239 $\pm$ 0.027  & 1.244 $\pm$ 0.008  & 1.247 $\pm$ 0.008  \\
\rowcolor{gray!25}
Cen A             & 1.226 $\pm$ 0.018  & 1.212 $\pm$ 0.014 & 1.218 $\pm$ 0.012  \\
Cen B             & 0.874 $\pm$ 0.028  & 0.859 $\pm$ 0.011 & 0.857 $\pm$ 0.009 \\
\rowcolor{gray!25}
Sun               & 0.998 $\pm$ 0.004 & 0.999 $\pm$ 0.004 & 0.998 $\pm$ 0.004  \\
\hline                                   
\end{tabular}
\label{radii_values}
\end{table*}
%
%
\begin{table*}
\centering 
\caption{
Inferred stellar masses from the different observable combinations. 
 }
\begin{tabular}{cccccccc}        
\hline 
Star          & Set~1 & Set~2 & Set~3 & Set~4 & Set~5\\
\hline\hline
Cyg A            & 1.05 $\pm$ 0.03 & 1.06 $\pm$ 0.01  & 1.06 $\pm$ 0.01 & 1.04 $\pm$ 0.03   & 1.06 $\pm$ 0.01  \\
\rowcolor{gray!25}
Cyg B            & 1.00 $\pm$ 0.03 & 1.01 $\pm$ 0.01 & 1.01 $\pm$ 0.01 & 0.98 $\pm$ 0.03     & 1.01 $\pm$ 0.01  \\
Doris             & 0.98 $\pm$ 0.04  & 0.94 $\pm$ 0.02  & 0.96 $\pm$ 0.01  & 0.96 $\pm$ 0.06 & 0.97 $\pm$ 0.01  \\
\rowcolor{gray!25}
Perky            & 1.09 $\pm$ 0.05 & 1.10 $\pm$ 0.02 & 1.11 $\pm$ 0.02 & 1.07 $\pm$ 0.05    & 1.11 $\pm$ 0.02  \\
Saxo2            & 1.15 $\pm$ 0.03  & 1.19 $\pm$ 0.02  & 1.19 $\pm$ 0.02 & 1.15 $\pm$ 0.03  & 1.19 $\pm$ 0.02  \\
\rowcolor{gray!25}
Cen A            & 1.10 $\pm$ 0.04  & 1.08 $\pm$ 0.04 & 1.10 $\pm$ 0.03 & 1.11 $\pm$ 0.04    & 1.08 $\pm$ 0.03 \\
Cen B            &  0.91 $\pm$ 0.03 &  0.89 $\pm$ 0.03 & 0.89 $\pm$ 0.03 & 0.91 $\pm$ 0.04   & 0.90 $\pm$ 0.03 \\
\rowcolor{gray!25}
Sun             & 0.99 $\pm$ 0.03 &  1.00 $\pm$ 0.01  & 1.00 $\pm$ 0.01 & 0.99 $\pm$ 0.03    & 1.00 $\pm$ 0.001 \\
\hline                                   
\end{tabular}
\label{masses_values}
\end{table*}
\subsection{Impact of a model-independent radius on the inferred mass}
\label{radius_results}
Table~\ref{constraints1} shows a combination of constraints used in examining the accuracy and precision of inferred masses when an interferometric radius is adopted instead of a stellar luminosity.
%
%
\begin{table}
\centering 
\caption{
Different combinations of seismic and atmospheric constraints.}
\begin{tabular}{cc}        
\hline 
Sets         & observational combinations \\
\hline\hline
4            &  [Fe/H], $T_{\rm{eff}}$, $R$ \\
\rowcolor{gray!25}
5            &  $\nu_i$, [Fe/H], $T_{\rm{eff}}$, $R$\\
\hline                                   
\end{tabular}
\label{constraints1}
\end{table}
%
\begin{figure}
    \includegraphics[width=\columnwidth]{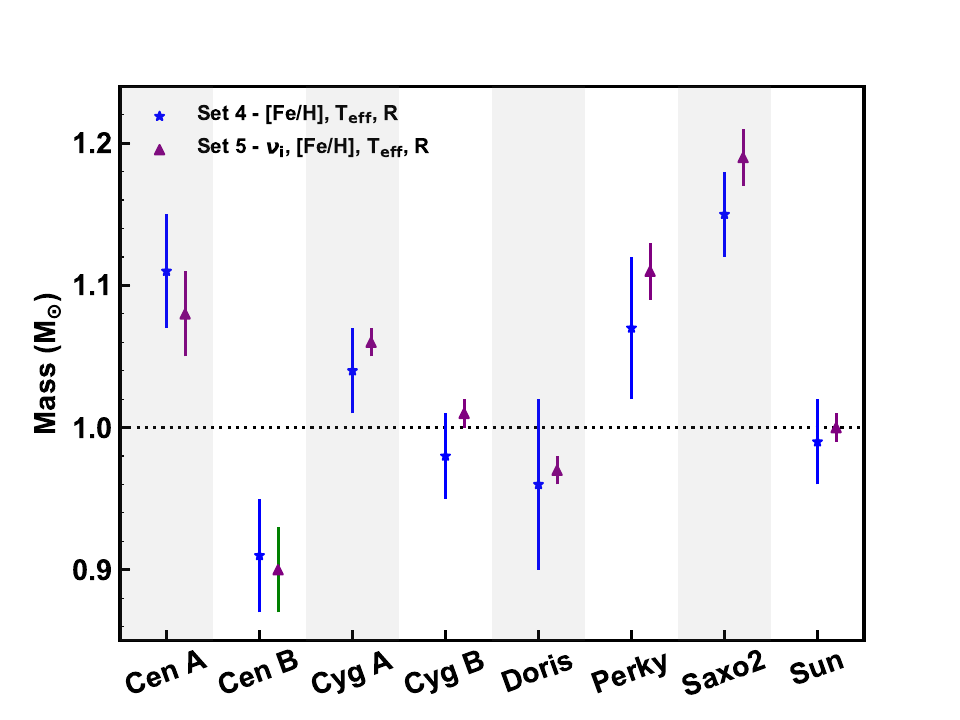}
    \includegraphics[width=\columnwidth]{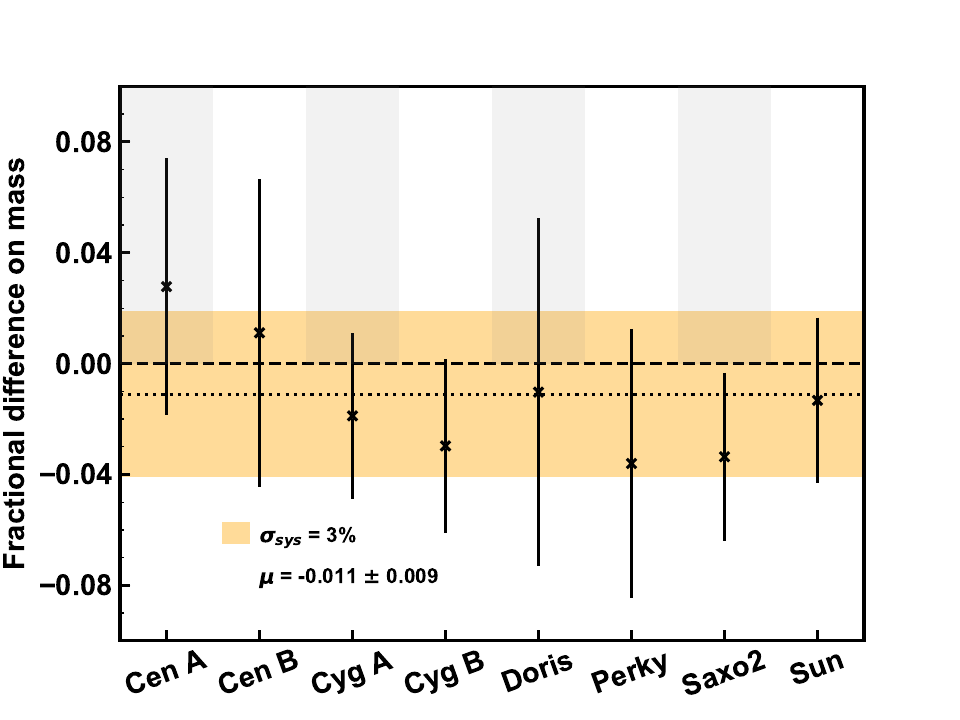}
      \caption{
      Top panel: comparison of the derived absolute masses and their associated uncertainties from different observable combinations with an interferometric radii taken into account in the optimisation process. Bottom panel: fractional difference in mass derived from Set~4 relative to Set~5.
      Light orange color shows the scatter (systematics) while the dotted black line shows an offset.
      } 
    \label{RandM}
\end{figure}
The top and bottom panels of Figure~\ref{RandM} show that the precision of the derived stellar masses improves when observable combinations are varied from Set~4 to Set~5, yielding a scatter of 3 per cent and an offset of  $-1.1 \pm 0.9$ per cent. It is also evident that the masses obtained using Set~4 are underestimated compared to those from Set~5 (See bottom panel of Figure~\ref{RandM}). It is important to note that Set~5 includes seismic data via the individual oscillation frequencies which places strong constraints on the stellar mean density, and with a known radius, the mass is indirectly constrained, while Set~4 does not place any constraints on the stellar mass. 
\begin{figure}
    \includegraphics[width=\columnwidth]{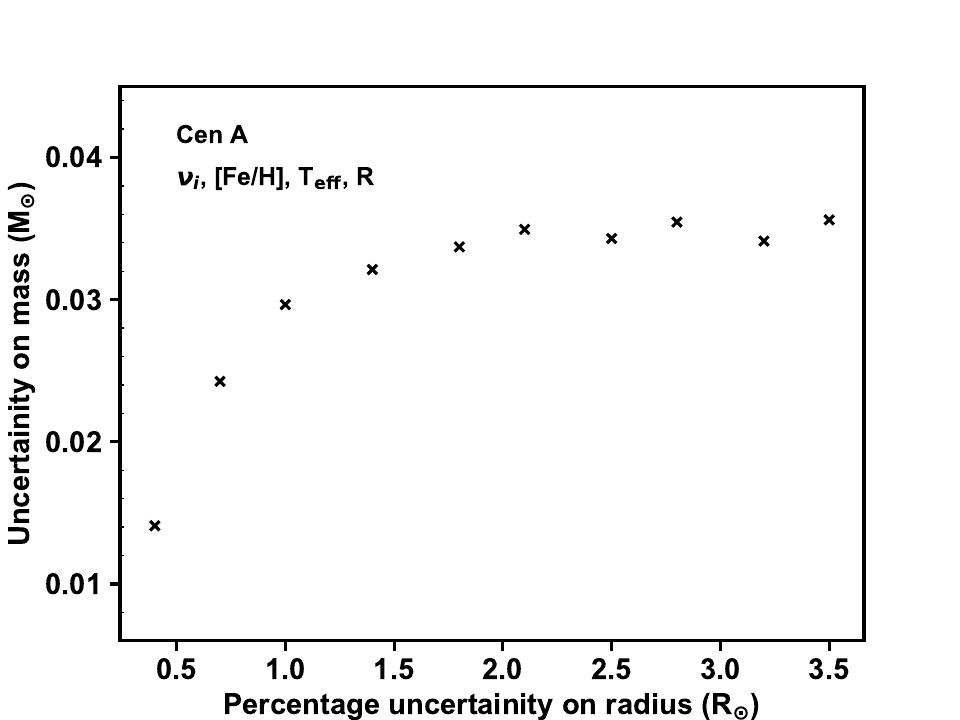}
    \includegraphics[width=\columnwidth]{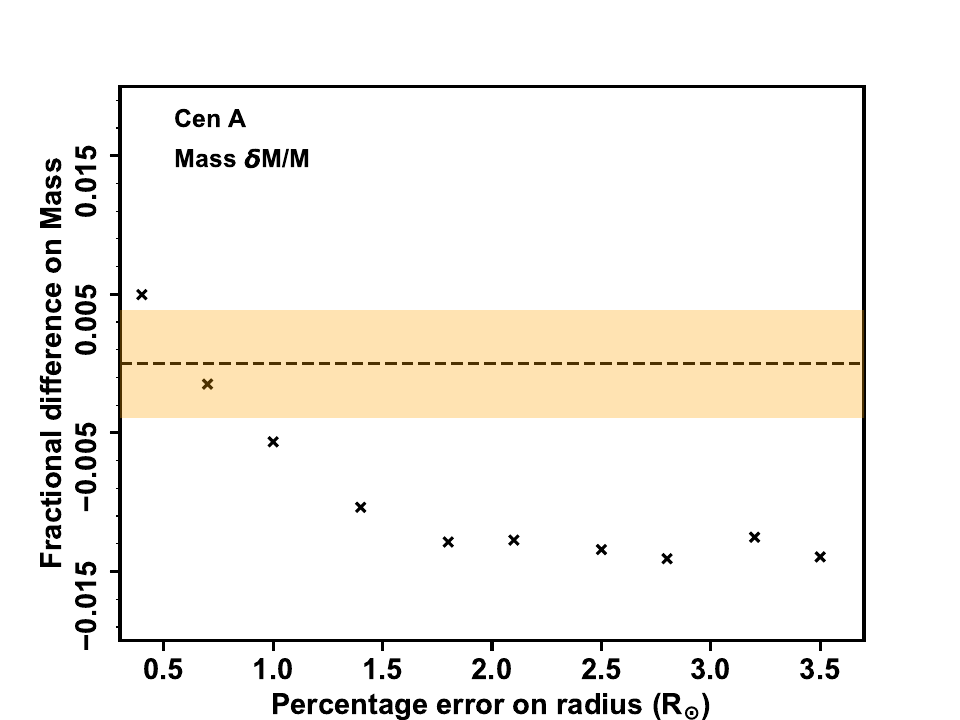}
      \caption{
      Statistical uncertainties (top panel) and fractional differences between the inferred masses and dynamical mass of $\alpha$~Centauri~A (bottom panel) against the varied uncertainties on interferometric radius. The orange region represents the dynamical mass range. The error on the interferometric radius was varied from 1$\sigma$ ($\sim$0.5 per cent) to 7$\sigma$ ($\sim$3.5 per cent).}
    \label{R_varied}
\end{figure}

Next, we assess how the precision of a model-independent radius (interferometric radius) impacts on the determination of a robust stellar mass. We consider $\alpha$~Centauri~A as a reference star for this exercise. $\alpha$~Centauri~A has both an interferometric radius and a dynamical mass with associated errors below 1 per cent, i.e.\ $R_{\rm{A}}$ = 1.2234 $\pm$ 0.0053 R$_\odot$ and $M_{\rm{A}}$ = 1.1055 $\pm$ 0.004 M$_\odot$, respectively \citep{2017Kervella}. We perform the best-fit model selection process by considering constraints in Set~5.
The error on the interferometric radius was varied from 1$\sigma$ to 7$\sigma$. For each combination, a mass is derived. The top panel of Figure~\ref{R_varied} highlights the trend of how the uncertainty on the inferred stellar mass varies with the uncertainty on the interferometric radius. For an interferometric radius with the uncertainty of $\lesssim$ 1 per cent, the uncertainties on the inferred masses are $\lesssim$ 2.5 per cent. In addition, an interferometric radius with the uncertainty of 0.5 per cent yields a corresponding uncertainty of about 1.5 per cent on stellar mass.
It is worth noting that the uncertainty on the inferred mass increases as that on the interferometric radius increases, and becomes relatively constant when the uncertainty on the interferometric radius is above $\sim$1.5 per cent (see top panel of Figure~\ref{R_varied}). At this level the interferometric radius is no longer restrictive, and the total mass uncertainty is determined by all the other measured quantities.

The bottom panel of Figure~\ref{R_varied} shows that the stellar mass is accurately determined when an interferometric radius with a precision below $\sim$1 per cent is used in the optimisation process. A model-independent radius at this level of precision dominates the seismic data in the determination of a stellar mass.
The impact of the interferometric radius vanishes once its uncertainty is above $\sim$1.5 per cent. This is because the seismic observables (individual oscillation frequencies) dominate other specified observables (including a non precise interferometric radius) towards the determination of the stellar mass. Our findings are consistent with predictions based on theoretical simulations by \citet{2007Creevey}. 
Based on these findings, in order to mitigate a scatter of about $\sim$5 per cent on the inferred asteroseismic stellar mass arising from variations in model physics and/or optimisation tool employed (e.g \citealt{2015Aguirr,2017Aguirre,2018MNRANsamba}), it would be vital to include an interferometric radius if available, among the observable constraints. However, we note that this is only relevant if the precision on the interferometric radius is about $\lesssim$ 1 per cent.

\subsection{Influence of the quality and length of seismic data on the inferred masses and radii.}
\label{reduced_models_seismic_results}
We examine how seismic data, when complemented with atmospheric constraints (effective temperature, metallicity, and luminosity),   performs towards yielding robust stellar masses and radii. 
This analysis is based on two stars, i.e 16~Cyg~A with high signal to noise observations from {\it{Kepler}} available  \citep{2017Lund} and $\alpha$~Centauri~A observed with ground-based observations, yielding oscillation frequencies with relatively low precision \citep{2010Meulenaer}. 
\begin{figure*}
    \includegraphics[width=\columnwidth]{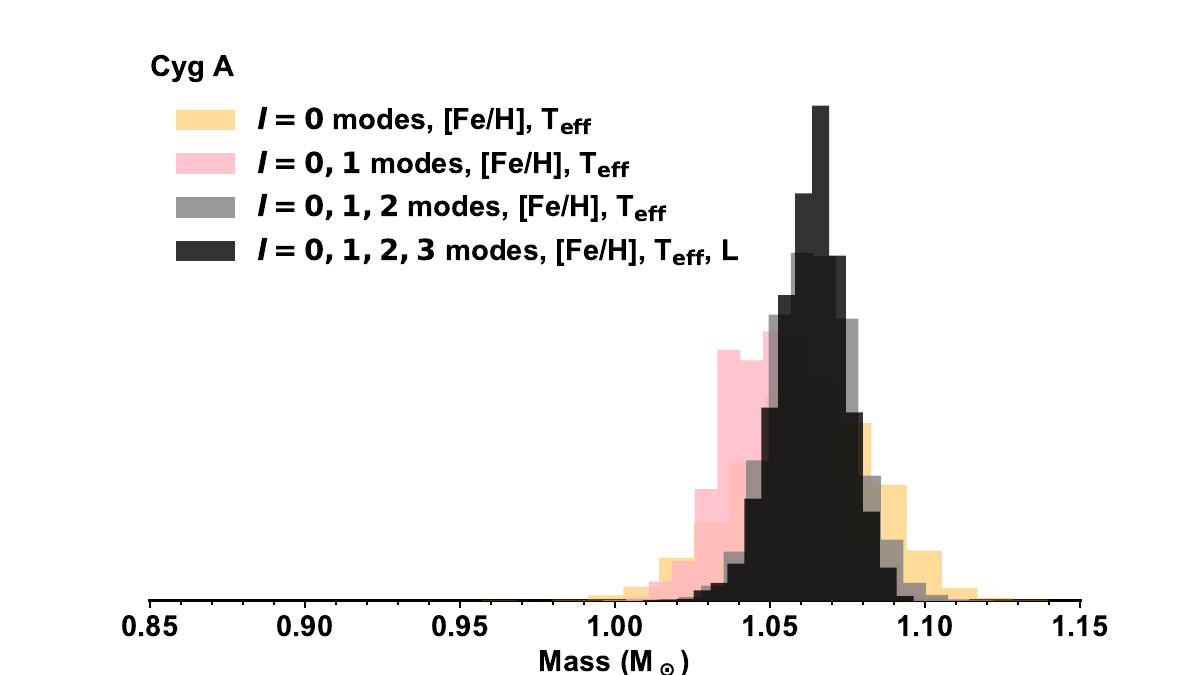}
    \includegraphics[width=\columnwidth]{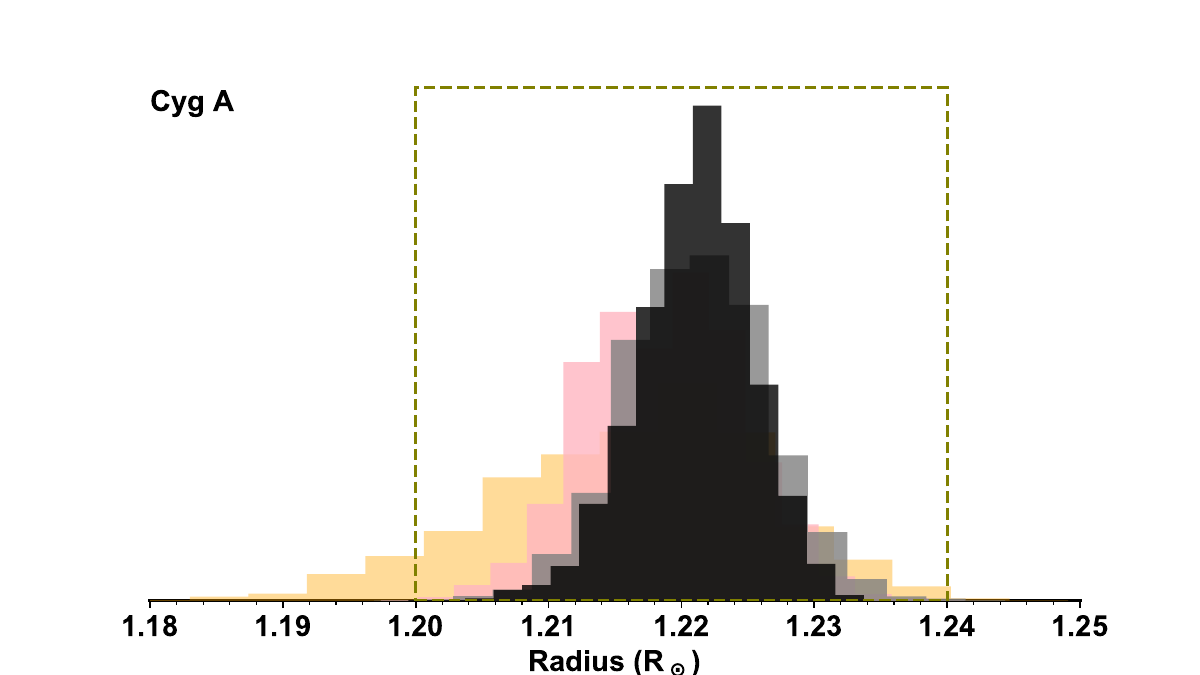}
    \includegraphics[width=\columnwidth]{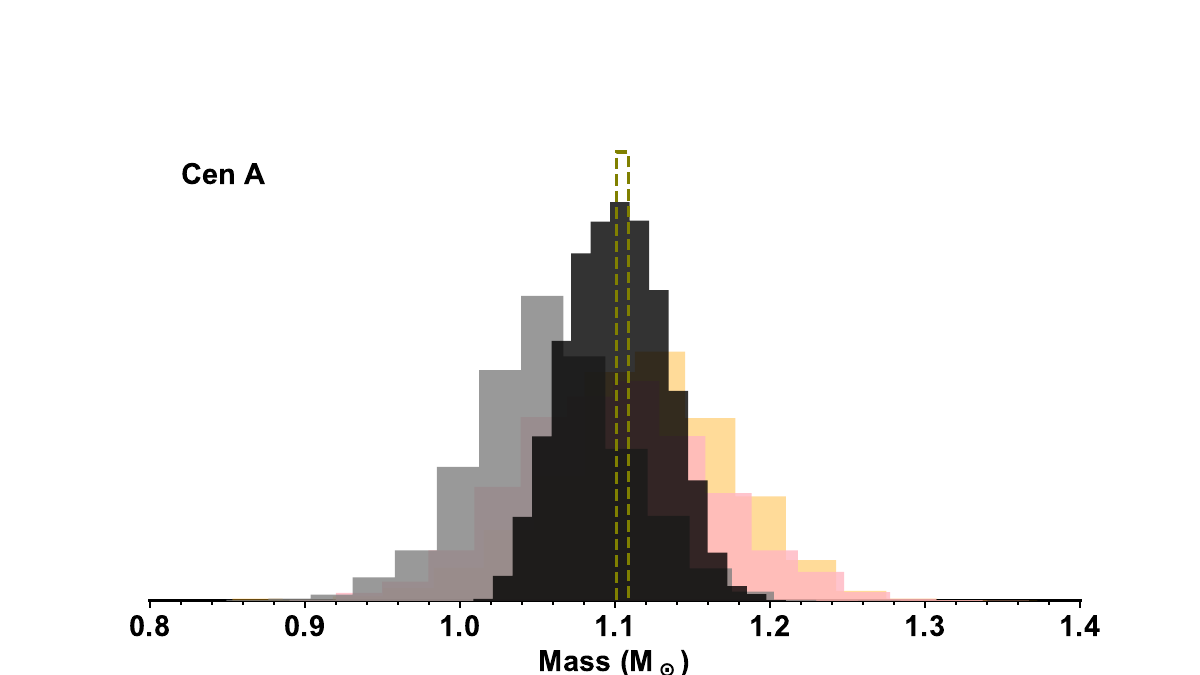}
    \includegraphics[width=\columnwidth]{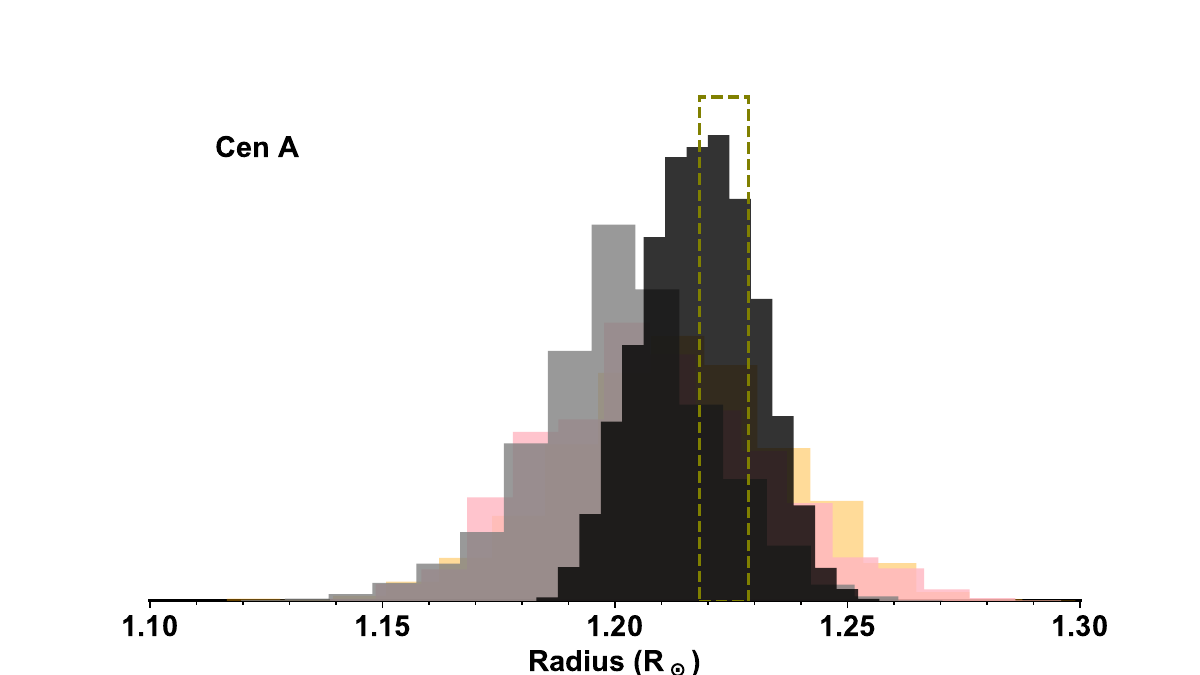}
      \caption{
      Normalised probability density distributions for mass (left panels) and radius (right panels) of 16~Cyg~A (top panels) and $\alpha$~Centauri~A (bottom panels). Color-coded according to the applied oscillation frequency modes and atmospheric constraints. The dashed olive lines correspond to independent radii and masses. 
      }
    \label{modes_only}
\end{figure*}
\begin{figure*}
    \includegraphics[width=\columnwidth]{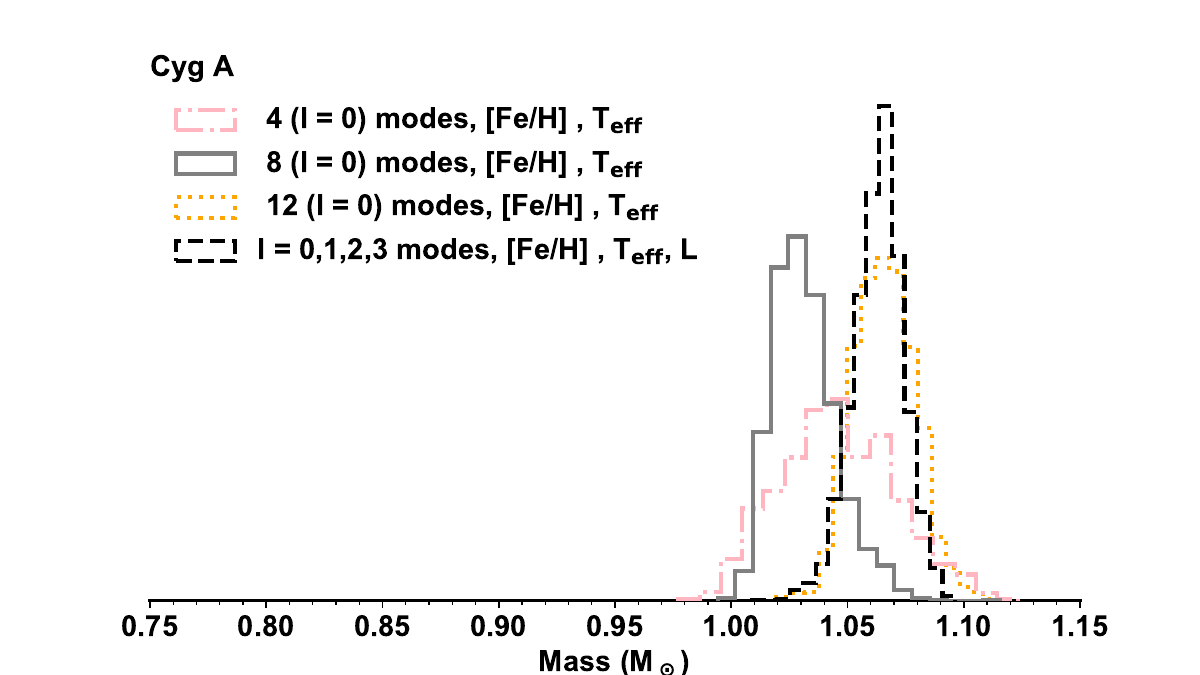}
    \includegraphics[width=\columnwidth]{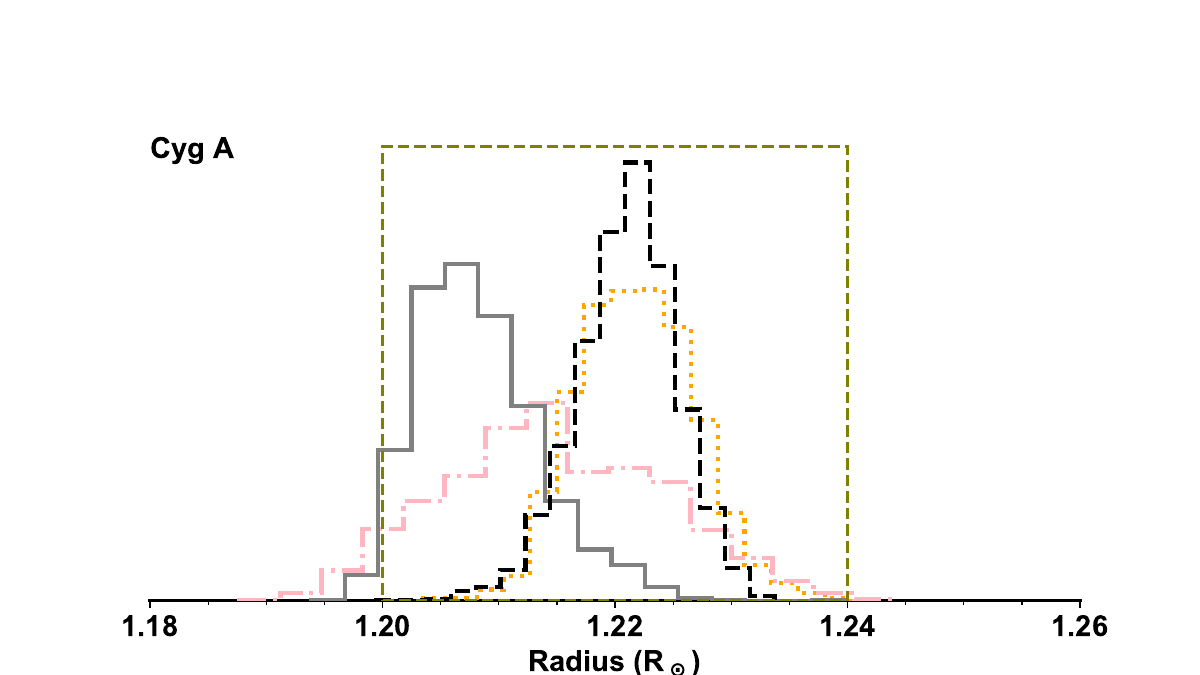}
    \includegraphics[width=\columnwidth]{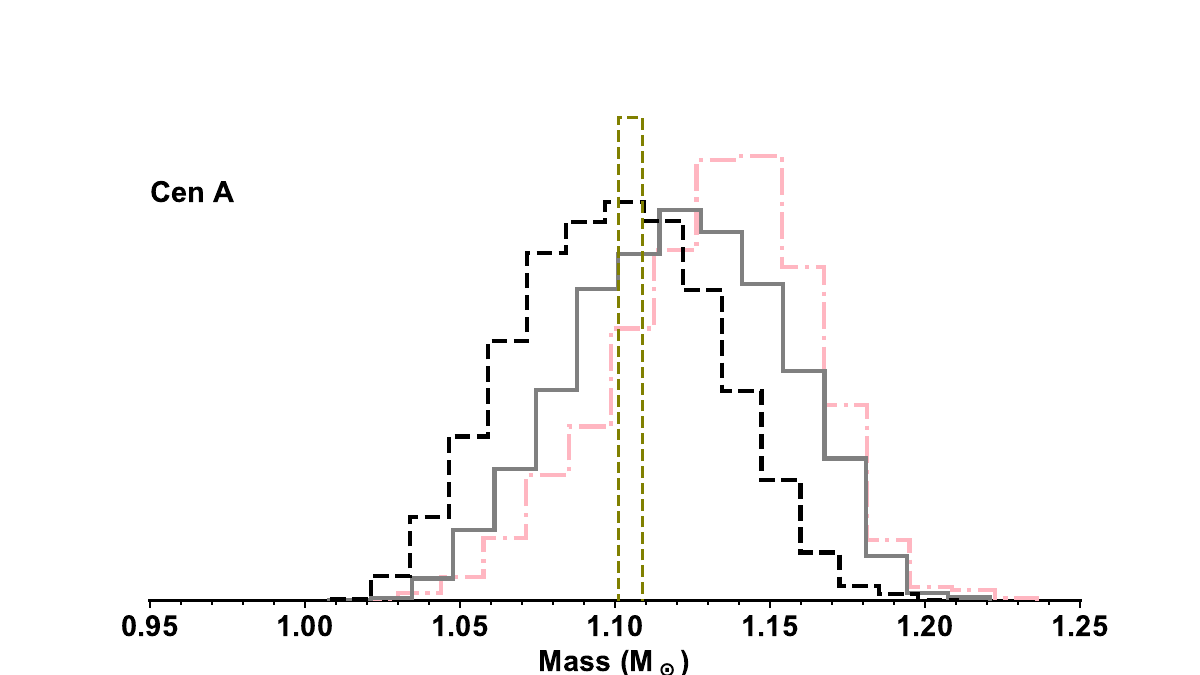}
    \includegraphics[width=\columnwidth]{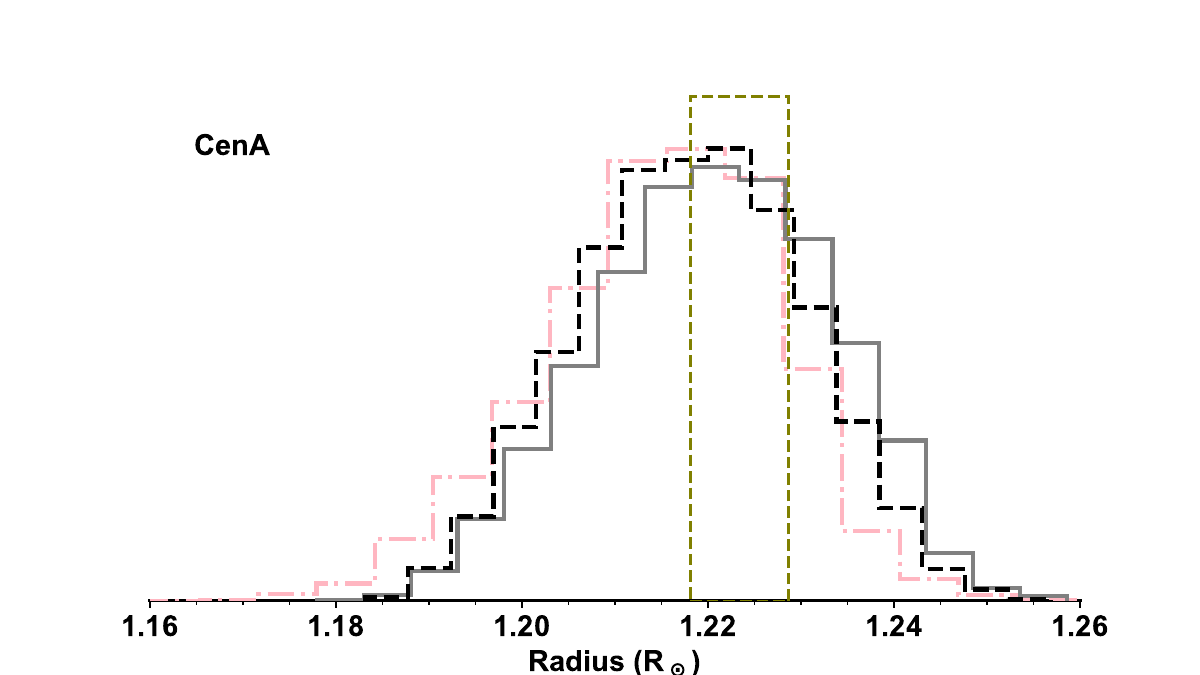}
      \caption{
      Normalised probability density distributions for mass (left panels) and radius (right panels) of 16~Cyg~A (top panels) and $\alpha$~Centauri~A (bottom panels). Color-code according to the number of frequency modes: Pink dashed dotted line - 4 ($l~=~0$) modes, Gray dashed line - 8 ($l~=~0$) modes, Orange dotted line - 12 ($l~=~0$) modes, and black dashed line - all ($l=~0,~1,~2,~3$) modes. The dashed olive lines correspond to independent radii and masses.
       }
    \label{mode_variations}
\end{figure*}
\begin{figure}
    \includegraphics[width=\columnwidth]{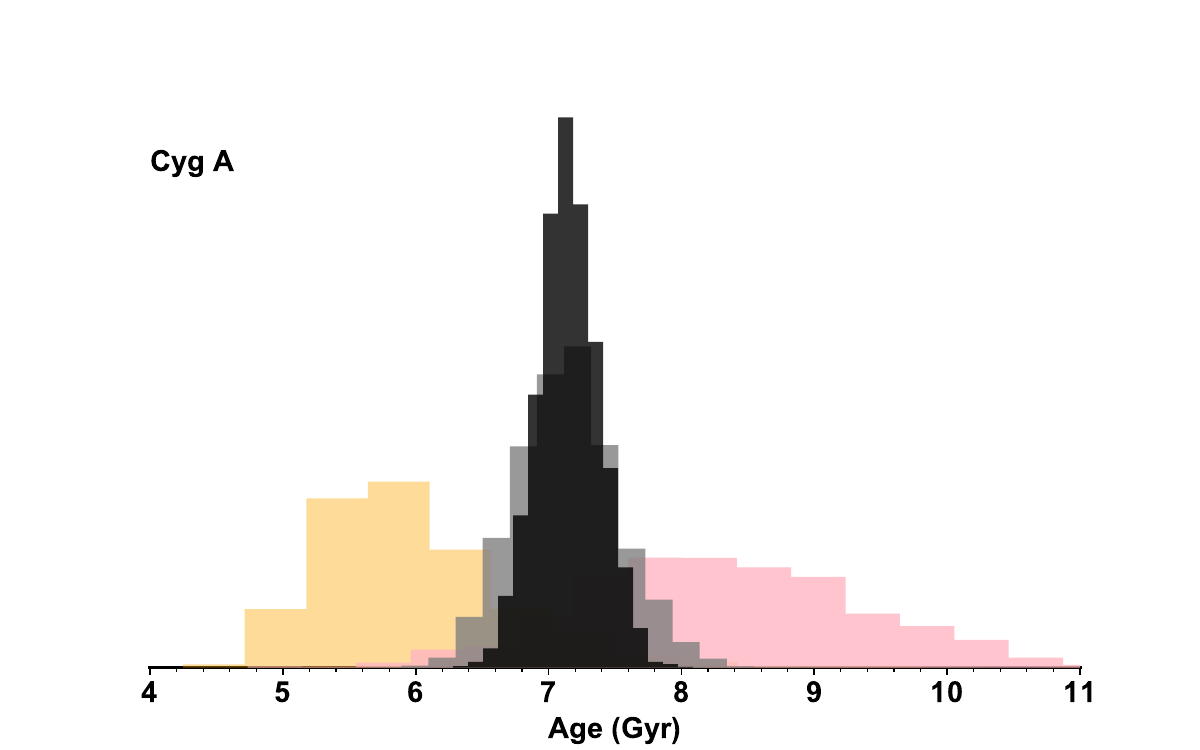}
    \includegraphics[width=\columnwidth]{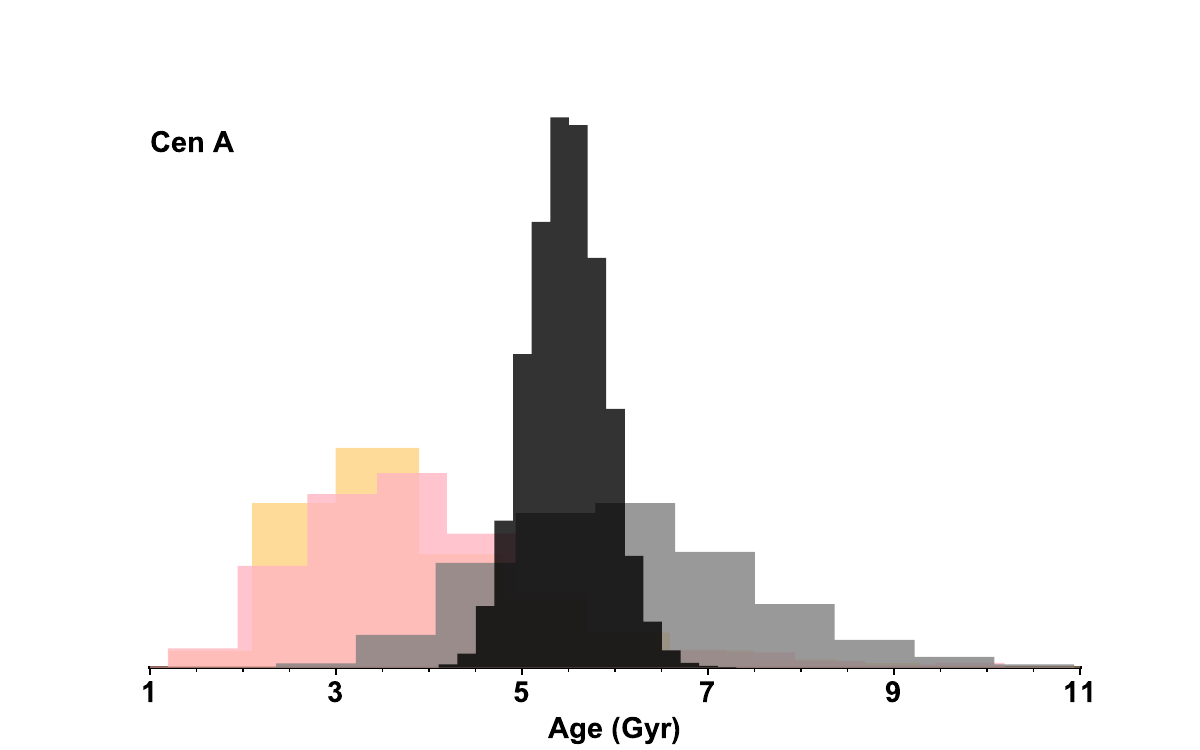}
      \caption{
      Same as Figure.~\ref{modes_only}, but for stellar age.
      }
    \label{Ages_modes_only}
\end{figure}
\begin{figure}
    \includegraphics[width=\columnwidth]{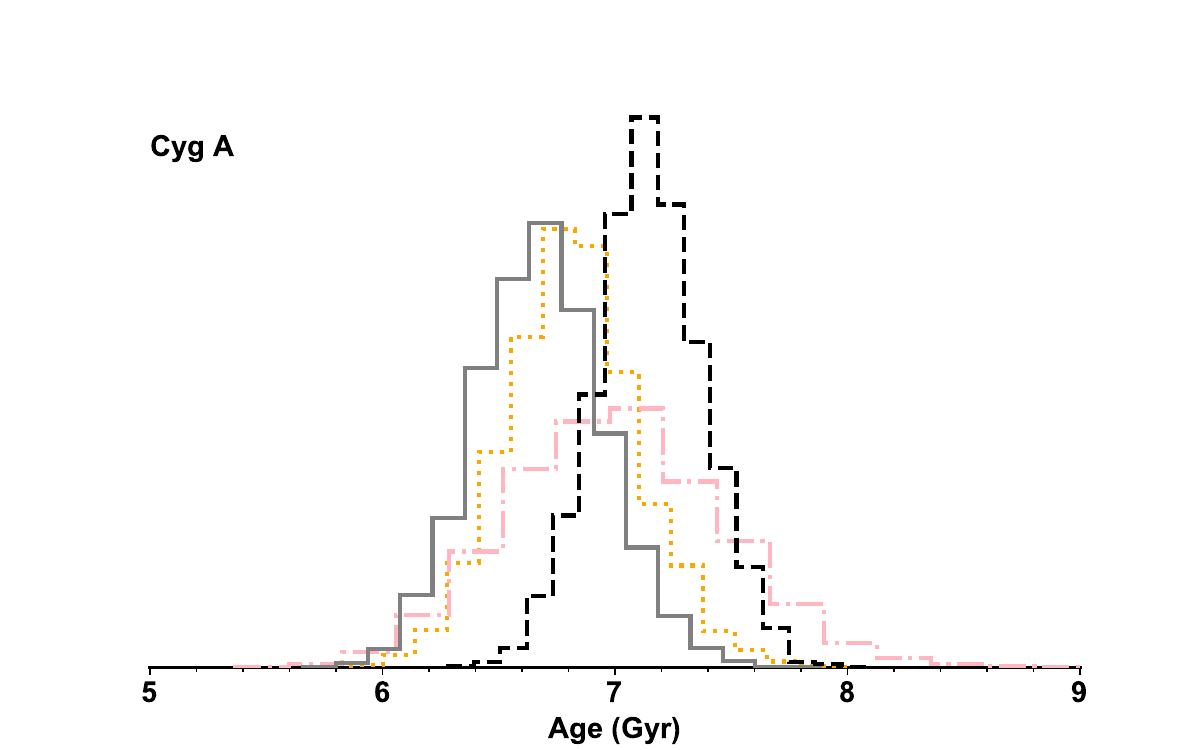}
    \includegraphics[width=\columnwidth]{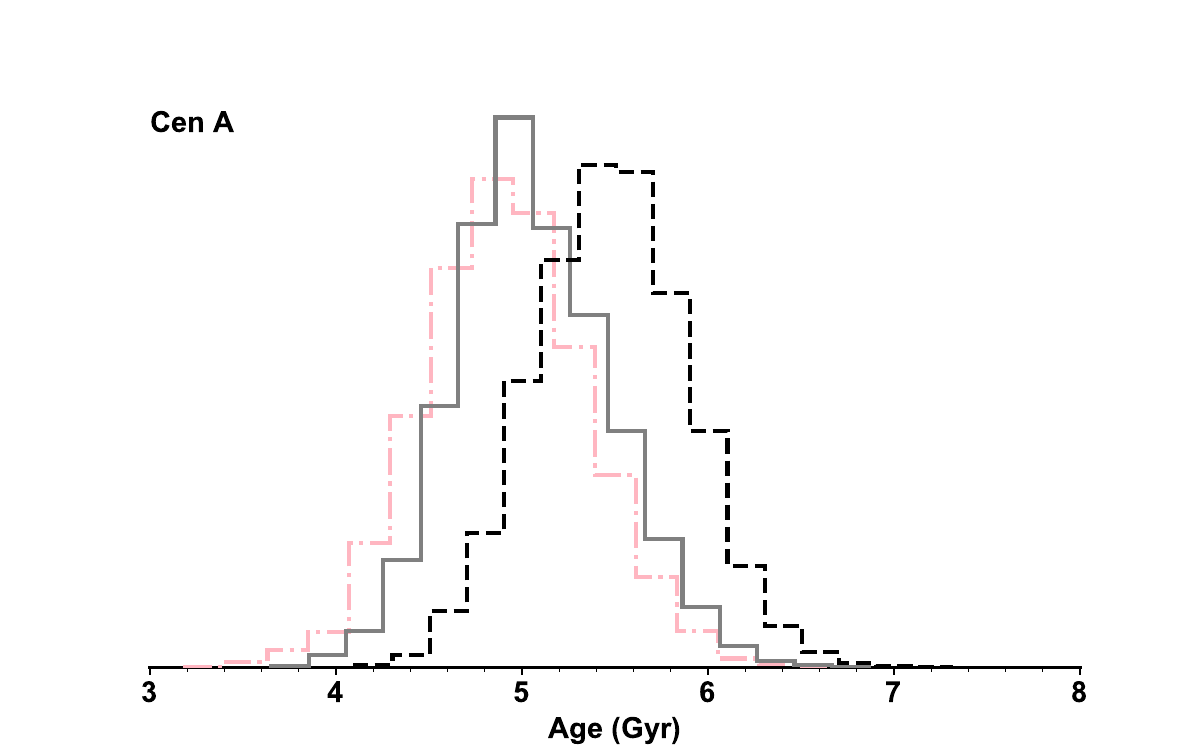}
      \caption{
      Same as Figure~\ref{mode_variations} but for stellar age.
      }
    \label{Ages_modes_only1}
\end{figure}
We perform different runs for both 16~Cyg~A and $\alpha$~Centauri~A fitting different seismic modes, i.e.\ all $l~=~0$ modes, $l~=~0, 1$ modes, $l~=~0, 1, 2$ modes, and $l~=~0, 1, 2, 3$ modes, complemented with atmospheric constraints. 
The top panels of Figure~\ref{modes_only} show that when the spherical degree modes are increased from $l~=~0$ to $l~=~0,~1,~2$ modes, the number of acceptable models reduces yielding more narrower probability distributions, thus demonstrating that more seismic information yields better constrained masses and radii measurements.  
In addition, the top right panel of Figure~\ref{modes_only} shows that the inferred radii from the different modes of 16~Cyg~A are also in excellent agreement with the interferometric radius. The bottom panels of Figure~\ref{modes_only} show that the optimal masses and radii of $\alpha$~Centauri~A, inferred when $l~=~0$ modes and $l~=~0,~1$ modes are used in the optimisation process, are in good agreement with the dynamical mass and interferometric radius, respectively. However, an offset in the derived mass and radius can be observed when modes of $l~=~0,~1,~2$ are used. This could partly stem from the available ground-based seismic data of $\alpha$~Centauri~A which is less precise. Complementing this seismic data with $l~=~3$ modes and atmospheric constraints with a Gaia-based luminosity, yields the optimal inferred mass and radius which are in good agreement with the dynamical mass and interferometric radius.

In Figure~\ref{mode_variations}, we explore the impact of varying the number of $l~=~0$ mode oscillation frequencies (i.e.\ from 4, 8, and 12 frequencies) on the inferred mass and radius. The top right panel of Figure~\ref{mode_variations}, shows that inferred radii of 16~Cyg~A obtained when the number of $l~=~0$ mode oscillation frequencies is varied are in good agreement with the interferometric radius. However, as the number of oscillation frequencies is increased (from 4, 8, and 12 frequencies), the number of acceptable models is reduced and narrower probability distributions are obtained. The top right panel of Figure~\ref{mode_variations} also shows that consistent probability distributions are obtained when 12 oscillation frequencies of $l~=~0$ mode  and  all $l~=~0,~1,~2,~3$ mode frequencies are used in the optimisation process, with the latter yielding a higher radius probability. Similar findings are obtained for the mass of 16~Cyg~A (see top left panel of Figure~\ref{mode_variations}). Unfortunately, no model-independent mass is available for this star, thus we ascertain the accuracy of the inferred masses by considering the case of $\alpha$~Centauri~A. 
When the number of $l~=~0$ mode oscillation frequencies is increased (from 4 to 8 and 12 frequencies), their dominance also increases over the atmospheric constraints (effective temperature, metallicity, and luminosity), leading to a more constrained/narrower probability distributions of both radius and mass (see top panels of Figure~\ref{mode_variations}). The bottom panels of Figure~\ref{mode_variations} show that the probability distributions of both mass (left bottom panel) and radius (right bottom panel) produced from the different $l~=~0$ mode oscillation frequencies are relatively consistent. This is because the seismic observations of $\alpha$~Centauri~A  are ground-based and not as precise as space observations (case of 16~Cyg~A). Thus little additional information is obtained from increasing the number of less precise $l~=~0$ mode oscillation frequencies which could play a significant role in further restricting the model selection process. From the bottom panels of Figure~\ref{mode_variations}, the inferred masses and radii from all the different distributions are in agreement within the uncertainties with the dynamical mass and interferometric radius, respectively. 
In general and based on the panels of Figure~\ref{mode_variations}, the precision and number of $l~=~0$ mode oscillation frequencies  play a vital role towards the determination of robust masses and radii in forward modelling processes.

Although this article focuses mainly on the inferred masses and radii, we briefly highlight the impact of the length and quality of seismic data on the inferred ages. We note that stellar ages inferred through asteroseismic forward modelling procedures are highly model-dependent and significantly vary depending on the model physics specified (e.g. \citealt{2015Aguirr,2017Aguirre,2018MNRANsamba}). Since no independent ages are available, our reference ages are based on those inferred using Set~3 (see Table~\ref{constraints}, i.e.\ from a combination of all individual oscillation frequencies of $l~=~0,~1,~2,~3$ modes, effective temperature, metallicity, and parallax-based luminosity). We again consider 16~Cyg~A and $\alpha$~Centauri~A because of the difference in the quality of the available seismic data. Figure~\ref{Ages_modes_only} shows that the probability distributions for both 16~Cyg~A and $\alpha$~Centauri~A span a wide range of values when only $l~=~0$ modes (orange)  and $l~=~0,~1$ modes (pink) are employed. This demonstrates that no strong constraint is placed on the age parameter. However, a reduced age parameter range (narrow probability distributions) is generated when $l~=~0,~1,~2$ modes are used. An excellent agreement can be seen in the top panel of Figure~\ref{Ages_modes_only} when probability distributions obtained using $l~=~0,~1,~2$ modes (gray) are compared with those obtained using constraints of Set~4 (black). The bottom panel of Figure~\ref{Ages_modes_only} shows that although $l~=~0,~1,~2$ modes and constraints in Set~4 yield consistent mean age value of $\alpha$~Centauri~A, the probability distributions are not as narrow as in the case of 16~Cyg~A (top panel of Figure~\ref{Ages_modes_only}). This may be attributed to the difference in the precision of the available seismic data (i.e.\ ground-based data for the case of $\alpha$~Centauri~A). Figure~\ref{Ages_modes_only1} shows that the consistent ages (agreement within 1$\sigma$) are obtained when the number of individual frequencies of only $l~=~0$ modes are varied (i.e.\ from 4, 8, and 12 frequencies). The top and bottom panels of Figure~\ref{Ages_modes_only1} show that increasing the number of $l~=~0$ modes frequencies from 4 to 8 leads to a higher age probability. No significant difference is observed when the number of $l~=~0$ modes frequencies is increased from 8 to 12. We note that only 10 ($l~=~0$ modes) frequencies are available for $\alpha$~Centauri~A, this explains why no probability distribution (orange) is shown  in the bottom panel of Figure~\ref{Ages_modes_only1}. 

\section{Summary and Conclusions}
\label{con}
In this article, we made use of a sample of main-sequence benchmark stars and explored the influence of a parallax-based luminosity when combined with other observational constraints (both seismic and atmospheric) towards improving the precision and accuracy of stellar mass and radius. Our selected stellar sample have model-independent radii available (interferometric radii), thus aiding to ascertaining not only the precision but also the accuracy of the derived radius parameters when considering different observational constraint combinations. Further, three stars in our stellar sample have model-independent masses allowing us to validate the accuracy of the inferred masses. In addition, we have explored the precision needed on a model-independent radius if it is to have a vital influence in the determination of a precise mass. Lastly, we explored how the quality and length of seismic data affects the robustness of inferred stellar masses and radii. Despite the fact that stellar ages are highly model-dependent, we briefly highlight the impact of the quality and length of seismic data on the inferred ages.
A comprehensive complementary study on stellar ages is presented in Kamulali et al. (in preperation).

%
 The precision of the inferred stellar mass improves when seismic and spectroscopic constraints (effective temperature and metallicity) are  complemented with a Gaia-based luminosity, i.e. with a scatter varying from 1.9 per cent to 0.8 per cent. 
 However, the inferred stellar radius is underestimated when compared to the interferometric radius, with an offset of $-1.9 \pm 0.7$ per cent and a scatter of up-to 1.9 per cent. 
Our findings also demonstrate that an independent radius with a precision below 1 per cent when applied in the optimisation process yields a mass with a precision below 1.5 per cent. This may hold the key in overcoming systematic uncertainties (scatter) induced on the inferred stellar mass arising from the improper description and modelling of stellar physics. Thus, these results recommend for an improvement in the interferometric measurements so at to reach a precision of $\sim$1 per cent in interferometric radius of solar-type stars, needed to infer robust masses.  

%
The results of this article also show that in cases where oscillation frequencies of higher degree modes (i.e.\ $l \gtrsim 1,2$ and $3$) are not available, robust masses and radii can still be attained with precise $l=0$ mode oscillations frequencies coupled with atmospheric constraints. However, the precision of the inferred masses and radii can be significantly improved when numerous precise $l=0$ mode oscillations frequencies ($>$ 8) are taken into account.
Our results also highlight that it is essential to complement non precise seismic data (ground-based) with atmospheric constraints (including a Gaia-based luminosity) if robust masses and radii are to be derived. 
We also note that individual oscillation frequencies of at least $l~=~0,~1,~2$ modes need to be taken into account in the forward modelling process if constrained age probability distributions are to be obtained, fewer modes result to the age probability distributions spanning a wide range of values.

This article demonstrates the relevance of {\it{\bf{{\it{Gaia}}}}}-based parallax measurements when complemented with other observational constraints in improving the precision and accuracy of inferred stellar masses and radii. The findings in this article should be taken into account in PLATO work packages related to the characterisation of stellar properties. Furthermore, an extended stellar sample with model-independent masses (dynamical masses) and with high quality seismic data made available from the PLATO mission will offer an opportunity to extensively examine the accuracy of masses derived through forward modelling, and the contribution of a stellar luminosity in yielding more accurate masses.

\section*{Acknowledgements}
\footnotesize
The authors acknowledge the anonymous referee  for  the  helpful  and  constructive remarks, which helped to significantly improve this paper.
B.N. acknowledges postdoctoral funding from the "Branco Weiss fellowship – Science
in Society" through the SEISMIC stellar interior physics group. JK acknowledges funding through the Max-Planck Partnership group - SEISMIC Max-Planck-Institut für Astrophysik (MPA) -- Germany and Kyambogo University (KyU) - Uganda. 
We also acknowledge funding from the UNESCO-TWAS programme, ``Seed Grant for African Principal Investigators'' financed by the German Ministry of Education and Research (BMBF).
BN thanks Nuno Moedas for providing evolution models for comparisons, 
Dr. Margarida S. Cunha and Dr. Savannah Nuwagaba for the useful discussions. 


\section*{DATA AVAILABILITY}
Stellar model data generated in this article using the stellar evolution code (MESA) will be made available on request. The MESA "inlists" describing the stellar physics highlighted in section~\ref{stellar_grid} will be provided. The oscillation frequencies data used for our sample stars is available in \citet{2017Lund}.


\bibliographystyle{mnras}
\bibliography{mybib} 






\bsp	
\label{lastpage}
\end{document}